\documentclass[aps,showpacs,nofootinbib]{revtex4}%
\usepackage[dvips]{graphicx}
\usepackage{amsmath}
\usepackage{amsfonts}
\usepackage{amssymb}%
\input epsf

\newcommand{\beq}{\begin{eqnarray}}
\newcommand{\eeq}{\end{eqnarray}}

\def\be{\begin{equation}}

\def\ee{\end{equation}}

\newcommand{\rem}[1]{}

\def\bit{\begin{itemize}}
\def\eit{\end{itemize}}

\def \be {\begin{equation}}
\def \ee {\end{equation}}
\def \bea {\begin{eqnarray}}
\def \eea {\end{eqnarray}}

\begin{document}


\begin{titlepage}
\hfill hep-ph/0412223 \break

\hfill \hfill UW/PT 04-12 \break

\hfill December 16, 2004 \break

\bibliographystyle{apsrev}
\preprint{UW/PT 04-12}

\title {In Search of Lonely Top Quarks at the Tevatron}
\author{Matthew T. Bowen}
\author{Stephen D. Ellis}
\author{Matthew J. Strassler}
\affiliation{Department of Physics,
P.O Box 351560, University of Washington,
Seattle, WA 98195}

\begin{abstract}
Single top-quark production, via weak-interaction processes, is an
important test of the standard model, potentially sensitive to new
physics. However, it is becoming known that this measurement is much
more challenging at the Tevatron than originally expected. We
reexamine this process and suggest new methods, using shape variables,
that can supplement the methods that have been discussed
previously. In particular, by focusing on correlations and
asymmetries, we can reduce backgrounds substantially without low
acceptance for the signal. Our method also allows for a
self-consistency check on the modeling of the
backgrounds. However, at the present time, serious systematic problems
remain, especially concerning the background from $W$-plus-jets; these
must be studied further by experimentalists and theorists alike.

\end{abstract}

\maketitle

\end{titlepage}

\section{Introduction}

The electroweak production of single top quarks is an important standard model
process which the Tevatron is guaranteed to be able to study. This reaction,
which has been investigated previously \cite{ssw,OTHERt}, is interesting both
because it provides a direct measurement of the $V_{tb}$ CKM\ element and
because it is sensitive to deviations of top quark physics from standard model
predictions, in particular, through their effects on the top-bottom-$W$ vertex
\cite{newsum}. Limits on single top production from Run I at the Tevatron have
been published \cite{cdfrunone,d0runone}, and the first Run II limits have
appeared \cite{cdfruntwo,d0runtwo}. In this article we discuss methods which
we hope will improve the significance of the measurement by using more
information encoded in the shape of the signal, in a way that will be less
sensitive to systematic errors than a simple counting experiment. However, we
also show that the size of the $W$-plus-jets
background, and the difficulty of predicting it accurately,
represent a serious obstacle.

\begin{figure}
[ptbh]
\begin{center}
\includegraphics[
height=1.497in,
width=4.3007in
]%
{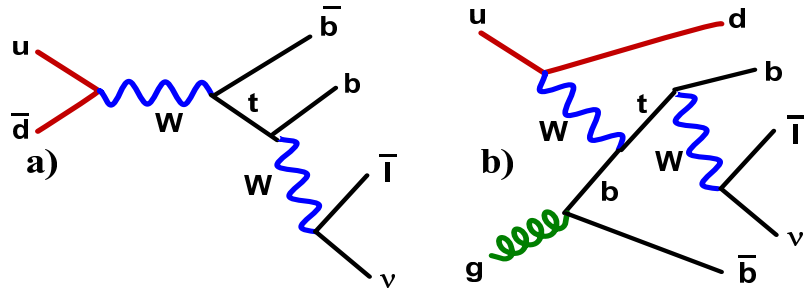}%
\caption{Single top quark production via (a) an $s$-channel $W$ boson; 
(b) a $t$-channel $W$ boson.}%
\label{tbboth}%
\end{center}
\end{figure}

Single top production is a very unusual process. At Fermilab energies, the
\textquotedblleft$tb$\textquotedblright\ production of a top quark and bottom
antiquark by an $s$-channel $W$ boson, as can occur through the diagram in
Fig.~\ref{tbboth}a,
has a lower cross-section than
\textquotedblleft$tbq$\textquotedblright
$t$-channel $W$ boson of a $t$, $\bar{b}$, and an extra light-quark
jet near the beam axis, as occurs through diagrams such as that in
Fig.~\ref{tbboth}b.\footnote{There is, as always, some ambiguity as to
whether the initial state contains a gluon which splits into a $b$ and
$\bar{b}$ as part of the scattering process or whether the initial
state contains a $b$ parton directly, \textit{i.e}., the splitting
process is part of the incoming wave function. Since the $\bar{b}$ jet
is generally not used in the analysis below, a careful examination of
this separation is not essential for us.}  The $tbq$ process has a
distinct shape, both because of its unusual initial state, the hard
light-quark jet (which has large $p_{T}$ and large pseudo-rapidity) in
the final state, and correlations between this jet and the lepton from
the top decay. In this paper, we will explore a method for using these
features to help separate single top from its major backgrounds:
$t\bar{t}$, $W$ plus jets, and QCD events.

This separation using the shape of the event is essential, because a simple
counting experiment is extremely difficult to carry out. Both our studies and
recent data indicate that the size of the relevant $W$-plus-jets background is
larger than anticipated a few years ago.\cite{ssw,OTHERt} This is compounded
by other issues, such as the 2\% decrease in Fermilab's energy compared to
expectations, and lower cross-section calculations for the signal
\cite{newcross,campbell,otherNLO}. We are concerned that systematic errors in
the understanding of the background will plague a direct counting experiment
at a level that will make any claims of discovery suspect. \ Thus, in our
view, additional methods, independent of (but perhaps to be used in
conjunction with) a counting experiment, are needed even for a discovery of
the process, as well as for a precision measurement. We will argue that it is
necessary, and possible, to reduce backgrounds further by using more
information about the final states. \ (Note that a shape fit using a single
variable was used in \cite{cdfrunone,cdfruntwo}.)

However, even with these improvements, our results suggest that the
measurement of single top remains challenging. We have found that there is no
simple way, even with 3 fb$^{-1}$ of integrated luminosity, to achieve both
good statistics and a satisfactory signal-to-background ratio. The essential
problem, compared to earlier and more optimistic studies \cite{ssw}, is that
the $W$-plus-jets sample with a single $b$-tagged jet is larger, less
predictable, and more variegated than was understood a few years ago. We will
discuss in some detail the difficulties with this background, and the need for
a wide range of efforts to bring it under control.

The organization of this paper is as follows. In Section II we discuss
the general structure of the events expected for both signal and
background events. Section III includes a detailed discussion of how
we have simulated both types of events, the cuts used to define the
event samples, the expected (and observed) differences between signal
and background event shapes, and new observables intended to highlight
these differences. In Section IV we address the essential issues of
uncertainties, both statistical and systematic, with a special focus
on the difficulties inherent in understanding the background arising
from $W$ boson production accompanied by jets. In the final section we
summarize our analysis and our conclusions.

\section{The Backgrounds to Single Top Production}

~From Fig.~\ref{tbboth}b 
one can see that a $tbq$ event has a final partonic state
consisting of at least the following: a charged lepton, a neutrino, $b$ and
$\bar{b}$ quarks, and a light quark. The bottom quark that comes directly from
the initial gluon tends to have small transverse momentum, and is rarely in
the pseudo-rapidity\ and $p_{T}$ range necessary to be identified by
$b$-tagging algorithms. Thus, in selecting $t$-channel signal events, one asks
for (a) one or more $b$-tagged jets, (b) significant missing energy (from the
neutrino), (c) one and only one isolated $e^{\pm}$ or $\mu^{\pm}$, (d) at
least one untagged jet. Typically, the highest-$p_{T}$ untagged
jet in a $t$-channel event is that from the light quark. Also, in typical
events a $t$ quark can be reconstructed from the tagged jet and a $W$, itself
reconstructed from the charged lepton and the missing energy. The kinematics
of the event tend to prefer a total visible scalar-summed transverse energy of
order $m_{t}$.

The $tb$ process has a $b$ quark jet and a $\bar b$ jet, along with a lepton
and a neutrino. In a significant number of events, one of the two $b$-jets
will not be tagged, so that the same criteria used for $tbq$ --- one
$b$-tagged jet, missing energy, a charged lepton and an untagged jet --- will
have moderately high efficiency for this process as well. Since the $tb$
process is smaller in cross-section and considerably less distinctive in shape
than $tbq$, and therefore harder to separate from background, it makes sense
for us to optimize our approach for $tbq$. We will not, in this paper, discuss
the measurement of the $tb$ and $tbq$ channels separately, as this will only
be possible well after the initial measurements of the combined production process.

The main backgrounds to single top production, which all can imitate the
signature just described, are \cite{ssw} (1) \textquotedblleft$t\bar{t}%
$\textquotedblright, top quark pair production, primarily from events where
only one of the top quarks decays leptonically, but also from events with two
leptonic decays; (2) \textquotedblleft QCD\textquotedblright\ events with 
fake missing energy and with either a fake
lepton  or a lepton from a heavy flavor decay; and, most
problematic, (3) \textquotedblleft$Wj^{n}$\textquotedblright\ events with a
leptonically-decaying $W$ plus some number $n\geq2$ of quark or gluon jets.

\subsection{Top pair production}

Top quark pair production is a formidable background to all single top
channels. After both top quarks decay, there are two high-$p_{T}$ $b$
quarks, which typically give rise to at least one $b$-tagged jet, and
two $W$ bosons which can decay hadronically or leptonically. The
signature for single top will be mimicked if only one $W$ decays
leptonically, or if both do and only one charged lepton is detected.
The present measurement of the cross-section for $t\bar{t}$ production
has a large uncertainty, though this uncertainty is expected to drop
to around 10\% \cite{ttxcross} by the end of Run II. However, even
this systematic uncertainty would prohibit observation of single top
above the $t\bar{t}$ background, so a substantial amount of $t\bar t$
must be cut away. The contribution of $t\bar t$ with one hadronically
decaying top is especially problematic, since often the
leptonically-decaying top quark can be reconstructed. On the other
hand, this background can be reduced using the fact that these events
tend to exhibit considerably more transverse energy than true single
top events, to have more jets, and to be more spherical.  These
kinematic handles are not present for events where both top quarks
decay leptonically, but such events are suppressed by both the
leptonic decay branching fraction and our requirement that only one
$e^{\pm}$ or $\mu^{\pm}$ be observed in the detector.  Furthermore,
the accurate reconstruction of a leptonically-decaying top quark is
more difficult in such events.

\subsection{QCD backgrounds}

Pure QCD events can give rise to fake leptons and provide heavy-flavor
jets (or jets which have no heavy-flavor but are mistagged.) As
fluctuations in energy measurement can lead to a fake missing $E_{T}$
signal, these QCD events form a background to single top
production. While the energetics tend to be lower than in signal
events, and the invariant mass of the ``lepton'', missing energy and
$b$-tagged jet exhibits no peak at $m_{t}$, the number of events is so
large that it is not obvious, without data (or a complete detector
simulation), whether this represents a relevant background. Recent
results from DZero \cite{d0runtwo} indicate that the number of QCD
events entering their single-top samples (defined by somewhat
different cuts than used here) is smaller than for other backgrounds,
and is, in first approximation, comparable in size to the single-top
signal. This conclusion depends in detail on cuts and on the flavor of
the lepton (muons being more prevalent than electrons in their
samples) and may differ for CDF.  As we will
see, the methods that we describe below provide substantial further
reduction of this QCD background and the systematic errors associated
with determining it, to the point that it should not pose a serious
issue. Consequently, we will largely disregard the QCD contribution to
the sample, except for a discussion in Sec.~III.D as to when and why this is
justified.

\subsection{$W$ plus jets}

The $W$-plus-jets background is much more challenging. The $Wj^{n}$ events
potentially entering the sample consist of a real $W$ boson decaying
leptonically, and at least two other quarks or gluons in the final state.
While the $Wj^{n}$ events do tend to have smaller energetics than single top,
and do not have a reconstructible $t$ quark, the number of events is so
large, and the energy resolution at Fermilab is sufficiently broad, that the
$Wj^{n}$ events form a large and problematic background to single-top
production. 

The difficulties involved in simulating $Wj^{n}$ events have been
discussed elsewhere \cite{method2,matrix} and we will not give a full
review, but it is important to note that there are special problems
for the sample with a single $b$-tag that neither untagged nor
double-tagged samples suffer from.  In particular, many different
partonic processes, with different shapes, contribute to the sample in
a fashion which is difficult to predict accurately.  We will discuss
this in detail in section IV.D.
As we will see, reducing the systematic error on the prediction and/or
measurement of this process is essential for success.

\section{Simulations, Cuts, Shapes and Methods}

\subsection{Simulation and Kinematic Cuts}

To model both signal and background, we have used MadEvent \cite{madevent} to
generate events, Pythia \cite{pythia} to then simulate showering and
hadronization (using the default value of the initial shower scale,
$\sqrt{\widehat{s}}$), and PGS \cite{pgs} to act as a fast detector
simulation. The single top $s$- and $t$-channel cross-sections have been
normalized to 0.88 pb and 1.98 pb respectively \cite{newcross}, and generated
with factorization and renormalization scales $\mu=m_{t}=175$ GeV. For the
$t\bar t$ process, the cross-section is normalized to 6.7 pb \cite{ttthcross}
and generated with factorization and renormalization scales also at $\mu
=m_{t}=175$ GeV.
For the $Wj^{n}$ channel, we have limited our simulations to $W$-plus-two-jets
(henceforth $Wjj$) at the MadEvent level, since we believe the uncertainties
in our simulations of this process are as large as the contribution at the
next order in $\alpha_{s}$. Samples for the $Wjj$ channel were generated using
a renormalization and factorization scale $\mu=M_{W}/2$, which is selected to
give the correct normalization at next-to-leading order \cite{MCFM}. We
employ cuts at the MadEvent level of $p_{T{j_{1}}}$, $p_{T{j_{2}}} >10$ GeV
(where $p_{T{j_{i}}}$ is the transverse momentum of the $i^{th}$ jet), with
angular separation $\Delta R\left( j_{1},j_{2}\right)  >0.4$, and $|\eta
_{j}|<4.0$.

The $b$-tagging is simulated as follows. Jets containing a $b$ quark (either
perturbatively or produced during showering) are taken to be tagged with an
efficiency of the form
$0.5\tanh\left(  p_{T}/36\ \text{GeV}\right) $, where $p_{T}$ is the
transverse momentum of the jet. Jets containing a $c$ quark (either
perturbatively or produced during showering) are taken to be tagged
with a rate of the form
$0.15\tanh\left(  p_{T}/42\ \text{GeV}\right)  $,
while jets containing no heavy flavor are taken to be mistagged
with a rate of the form
$0.01\tanh\left(  p_{T}/80\ \text{GeV}\right)  $.\footnote{There is no
agreed-upon convention for the light-quark/gluon mistagging function. The
$\tanh$ form of the mistagging function is \textit{not} what is used in the
default PGS detector simulation, which instead is $-5.54\times10^{-5} +
1.66\times10^{-7}(p_{T}/1 \mathrm{\ GeV})^{2.507}$. Based on the fact that
this function leads to large mistagging rates at very high $p_{T}$, in
contradiction to measurements at CDF \cite{cdftagrates}, we have changed this
mistagging function within PGS to a $\tanh$ form. The overall mistagging rate
that we use is larger than assumed in other papers, including \cite{ssw} and
the recent work of \cite{campbell}, where the size of $Wj^{n}$ backgrounds is
smaller as a result. The true form of the mistagging function is detector- and
algorithm-dependent, and different mistagging functions do change the shape of
certain distributions. Our results suggest that uncertainties in this function
lead to important, but not dominant, systematic uncertainties in the $Wj^{n}$
background.}

 From the events generated in this fashion, we define our initial
event sample to be those events that contain \textit{one and only one}
isolated charged lepton, missing energy, \textit{one or more}
$b$-tagged jets, and \textit{one or more} untagged jets. These objects
satisfy the ``basic'' cuts listed in Table \ref{cuts1}.
The $p_{T}$ constraints in this and later tables
apply only to the highest-$p_{T}$ $b$-tagged jet and the
highest-$p_{T}$ untagged jet.
Additional cuts are necessary in order to bring backgrounds down to
a reasonable level.

Given that $Wj^{n}$ tends, compared to the signal, to have lower energy and
fewer jets, and that $t\bar{t}$ tends to have higher energy and more jets,
there are two natural choices of variables to cut on that take advantage of
the overall kinematics of the events without appealing to event shapes. One
possibility is to cut on the total transverse energy of the event; a second
would be to cut on the number of high-$p_{T}$ jets. The question of which of
these (or whether a combination thereof) has the lowest theoretical
uncertainties has not been resolved and we will not attempt to answer this
very important question definitively here.  For this study we have chosen to
cut on the quantity
\[
H_{T}=\sum_{\mathrm{jets}}(p_{T})_{i}+(p_{T})_{\ell}+\slash\hskip-.11inE_{T}%
\ ,
\]
where the sum is over all jets with $p_{T}>20$ GeV and $|\eta|<3.5$,
$(p_{T})_{i}$ is the magnitude of the transverse momentum of the $i^{th}$ jet,
$(p_{T})_{\ell}$ that of the lepton, and $\slash\hskip-.11inE_{T}$ is the
missing transverse energy in the event. Large values of $H_{T}$ tend to favor
$t\bar{t}$ final states, while lower values favor $Wj^{n}$ final states, with
the signal contribution peaking at intermediate values.

Furthermore, since the signal involves a $t$-quark, we also impose a
requirement that the invariant mass of the lepton, neutrino, and the leading
tagged jet be approximately equal to the top quark mass. In doing so, we must
reconstruct the neutrino's momentum component $p_{\nu,z}$ along the beam axis,
which has an ambiguity. We require that $(p_{\ell}+p_{\nu})^{2}=m_{W}^{2}$,
and among the two solutions for $p_{\nu,z}$ we choose the solution with
smallest absolute magnitude. (For complex solutions, only the real part is used.)

As suggested earlier, we find that such cuts cannot decrease the backgrounds
to the point that they are comparable to the signal. Two choices of
\textquotedblleft intermediate\textquotedblright\ and \textquotedblleft
hard\textquotedblright\ cuts are indicated in Tables \ref{cuts2i} and
\ref{cuts2h}. The resulting numbers of expected events for an integrated
luminosity of 3 fb$^{-1}$, summing over $e^{\pm}$ and $\mu^{\pm}$ (and thus
including both $t$ and $\overline{t}$), can be seen in Table \ref{results1}.
We show the number of events which survive both the basic cuts of Table
\ref{cuts1}, the intermediate cuts of Table \ref{cuts2i}, and the hard cuts of
Table \ref{cuts2h}. Consistent with \cite{ssw}, we find that while all of the
cuts contribute to the background reduction, the $Wj^{n}$ channel is reduced
primarily by a combination of the stiffer $p_{T}$ cuts and the
\textquotedblleft$m_{t} $\textquotedblright\ cut, while the $t\overline{t}$
background is affected primarily by the upper $H_{T}$ cut.\ While the basic
cuts reveal a signal to background ratio of approximately 1:21, this improves
to 1:7.4\ and 1:4.9\ using the intermediate and hard cuts. This is, at best, disappointing.

\begin{table}[ptb]
\medskip
\par
\begin{center}%
\begin{tabular}
[c]{|c|c|c|}\hline\hline
\multicolumn{1}{||c|}{Item} & \multicolumn{1}{||c|}{$p_{T}$} &
\multicolumn{1}{||c||}{$\left\vert \eta\right\vert $}\\\hline\hline
$\ell^{\pm}$ & $\geq15$ GeV & $\leq2$\\\hline
MET $\left(  \nu\right)  $ & $\geq15$ GeV & -\\\hline
jet $\left(  b\text{-tag}\right)  $ & $\geq20$ GeV & $\leq2$\\\hline
jet $\left(  \text{no }b\text{-tag}\right)  $ & $\geq20$ GeV & $\leq
3.5$\\\hline
\end{tabular}
\end{center}
\caption{Basic cuts for initial sample. Here $p_{T}$ is the transverse
momentum and $\eta$ is the pseudo-rapidity.}%
\label{cuts1}%
\end{table}

\begin{table}[ptb]
\medskip
\par
\begin{center}%
\begin{tabular}
[c]{|c|c|c|}\hline\hline
\multicolumn{1}{||c|}{Item} & \multicolumn{1}{||c|}{$p_{T}$} &
\multicolumn{1}{||c||}{$\left\vert \eta\right\vert $}\\\hline\hline
$\ell^{\pm}$ & $\geq15$ GeV & $\leq2$\\\hline
MET $\left(  \nu\right)  $ & $\geq15$ GeV & -\\\hline
jet $\left(  b\text{-tag}\right)  $ & $\geq20$ GeV & $\leq2$\\\hline
jet $\left(  \text{no }b\text{-tag}\right)  $ & $\geq20$ GeV & $\leq
3.5$\\\hline\hline
& Min & Max\\\hline
$H_{T}$ & $\geq180$ GeV & $\leq250$ GeV\\\hline
\textquotedblleft$m_{t}$\textquotedblright & $\geq160$ GeV & $\leq190$
GeV\\\hline
\end{tabular}
\end{center}
\caption{Representative \textquotedblleft intermediate\textquotedblright%
\ cuts.}%
\label{cuts2i}%
\end{table}\begin{table}[ptbptb]
\medskip
\par
\begin{center}%
\begin{tabular}
[c]{|c|c|c|}\hline\hline
\multicolumn{1}{||c|}{Item} & \multicolumn{1}{||c|}{$p_{T}$} &
\multicolumn{1}{||c||}{$\left\vert \eta\right\vert $}\\\hline\hline
$\ell^{\pm}$ & $\geq15$ GeV & $\leq2$\\\hline
MET $\left(  \nu\right)  $ & $\geq15$ GeV & -\\\hline
jet $\left(  b\text{-tag}\right)  $ & $\geq60$ GeV & $\leq2$\\\hline
jet $\left(  \text{no }b\text{-tag}\right)  $ & $\geq30$ GeV & $\leq
3.5$\\\hline\hline
& Min & Max\\\hline
$H_{T}$ & $\geq180$ GeV & $\leq250$ GeV\\\hline
\textquotedblleft$m_{t}$\textquotedblright & $\geq160$ GeV & $\leq190$
GeV\\\hline
\end{tabular}
\end{center}
\caption{Representative \textquotedblleft hard\textquotedblright\ cuts.}%
\label{cuts2h}%
\end{table}\begin{table}[ptbptbptb]
\medskip
\par
\begin{center}%
\begin{tabular}
[c]{|c|c|c|c|}\hline\hline
\multicolumn{1}{||c|}{Channel} & \multicolumn{1}{||c|}{Basic Cuts} &
\multicolumn{1}{||c|}{\textquotedblleft Intermediate \textquotedblright\ Cuts}
& \multicolumn{1}{||c||}{\textquotedblleft Hard\textquotedblright%
\ Cuts}\\\hline\hline
$tbq$ & $298$ & $67$ & $30$\\\hline
$tb$ & $145$ & $27$ & $13$\\\hline
$t\bar{t}$ & $2623$ & $140$ & $57$\\\hline
$Wjj$ & $6816$ & $550$ & $152$\\\hline\hline
$\left(  tbq+tb\right)  /\left(  t\bar{t}+Wjj\right)  $ & $0.047$ & $0.14$ &
$0.21$\\\hline
\end{tabular}
\end{center}
\caption{Numbers of events for 3 fb$^{-1}$ (summed over $t$ and $\bar{t}$, $e
$ and $\mu$ channels) for the three sets of cuts in Tables \ref{cuts1}%
--\ref{cuts2h}.}%
\label{results1}%
\end{table}

The difference in the results between our study and that of \cite{ssw}
is striking, and requires an explanation.\footnote{Due to our
different cuts and somewhat different approaches, a detailed
comparison between the results of the two analyses is difficult. To
obtain some quantitative sense of the differences let us focus on the
ratio of signal ($tbq+tb$) to $Wjj$ background, which is where the
bulk of the difference arises and which provides an upper limit on the
full signal to background ratio. For example, consider this ratio for
the ``basic cuts'' results in Table \ref{results1} yielding a ratio of
0.065. This is most usefully compared with the middle column (without
parentheses) in Table 3 of \cite{ssw}, where the corresponding ratio
is 0.24. The factor of nearly 4 difference results primarily from our
much larger estimate of the $Wjj$ contribution (larger by more than a
factor of 3). Adding the $m_{t}$ cut in \cite{ssw} increases their
value for the ratio to 0.77. However, with our much larger $Wjj$
contribution, we can improve this ratio to only 0.28 when using the
``hard'' cuts that include a cut on $m_{t}$, \textit{i.e.}, the factor
of 3 larger $Wjj$ background is still there. The jet veto used in
\cite{ssw} (the last column in their Table 3) only reduces the
background from $t\bar t$ (and also the signal), and does not help
reduce the $Wjj$ background.} We believe the main effects can be
accounted for straightforwardly. First, in the present study we have
the benefit of recent next-to-leading-order calculations \cite{MCFM},
which increase the overall rate for $Wj^{n}$ by of order 50 \%
compared to that used in \cite{ssw}. (In our leading-order
computations, this is effectively obtained through our lower choice of
renormalization scale.) Second, we find a much larger number of tagged
jets in the $Wj^{n}$ channel, because we include (by simulating parton
showering) the fragmentation of leading-order partonic gluons into
heavy flavor jets at subleading orders. (This effect would appear
already in a next-to-leading-order calculation, such as performed
recently in \cite{campbell}.) Third, our more pessimistic estimate of
energy resolution at the Fermilab detectors forces us to use a wider
$m_{t}$ window cut in order to have sufficient acceptance for the
signal; this lets in more $Wj^{n}$ background. Fourth, we use a more
pessimistic $b$-tagging rate (50\% \textit{vs} 60\%) and
light-quark-jet mistagging rate (1.0\% \textit{vs} 0.5\%), a more
pessimistic charm-to-bottom ratio in tagging (1:3.3 \textit{vs}
1:4.0), and more realistic $p_{T}$ distributions in tagging functions;
these all hurt the signal-to-background ratio and the efficiency for
the signal.  Other small negative effects include a lower center of
mass energy (1.96 TeV \textit{vs.} 2.00 TeV), and a lower
cross-section for the signal (due largely to a change in parton
distribution functions.) Note also that we have used $m_{t}=175$ GeV,
so our results may even be slightly optimistic in this regard.

\begin{figure}[ptb]
\begin{center}
\includegraphics[
height=4.3016in,
width=5.5754in
]{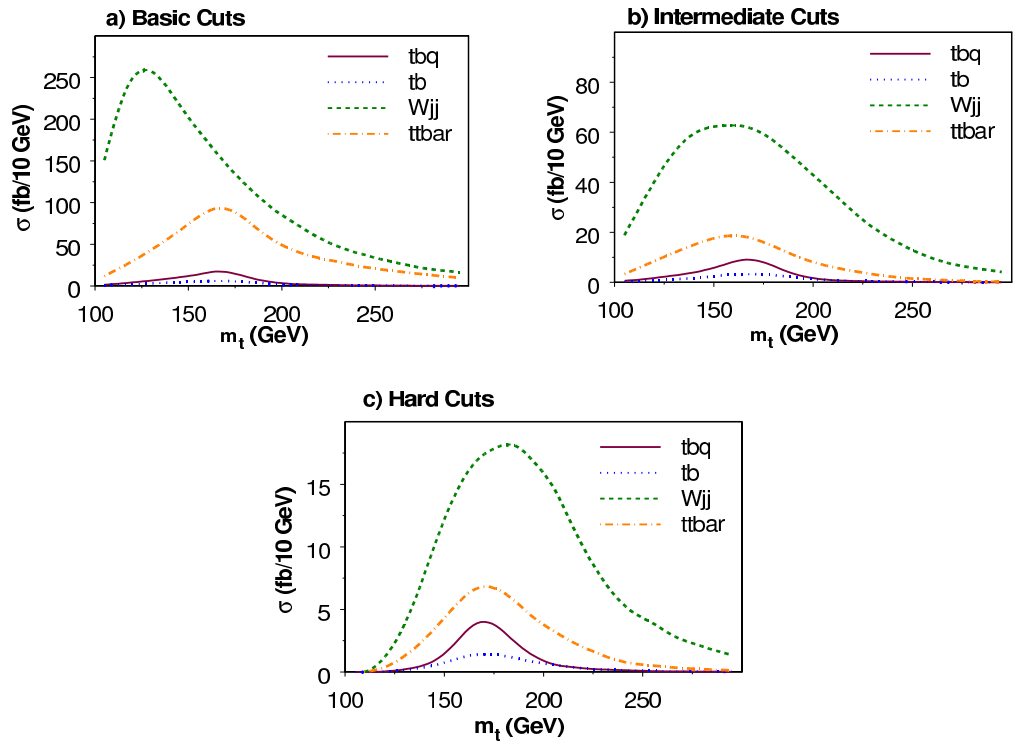}
\end{center}
\caption{Distribution of the reconstructed ``$m_{t}$'', the invariant mass of
the charged lepton, neutrino and highest-$p_{T}$ $b$-tagged jet, for the
backgrounds and signals. Additional details concerning the construction of
this observable are given in the text. The three figures show the results for
the (a) basic, (b) intermediate and (c) hard cuts given in Tables \ref{cuts1}%
-\ref{cuts2h}, with the cut on ``$m_{t}$'' omitted. }%
\label{fig:mtdistrib}%
\end{figure}

The numbers in Table \ref{results1} suggest that a further factor of
3--5 improvement in the signal-to-background ratio via more aggressive
cuts will come at the cost of a factor of 5--10 reduction in the
signal accompanied by a factor of 15--45 reduction in the
background. The essential question then is whether a reduction by such
a large factor can be achieved without large systematic and
theoretical errors. From Table \ref{results1}, one can see that
systematic errors below about 10\% in $Wj^{n}$ are needed for a
discovery.  Unfortunately, the method for removing the background
suggested in \cite{ssw}, namely, to use a jet veto to reduce $t\bar t$
to a small contribution, and do a sideband analysis on either side of
the $m_{t}$ window cut to remove $Wj^{n}$, is unlikely to work with
such a large $Wj^{n}$ background. This is illustrated in
Fig.~\ref{fig:mtdistrib}. With basic cuts, the $Wj^{n}$ background
(whose ``$m_{t}$'' distribution falls steeply and monotonically above
$125$ GeV) is very large, as shown in Fig.~\ref{fig:mtdistrib}a.  A
sideband analysis with a window centered around $175$ GeV would be
subject to large statistical errors. Meanwhile,
the intermediate and hard cuts, shown in Figs.~\ref{fig:mtdistrib}b
and \ref{fig:mtdistrib}c, tend to shape the $Wj^{n}$ events such
that the $Wj^{n}$ background is no longer monotonic near and/or across
the $m_{t}$ window, making a sideband analysis problematic. Thus the
shape of the ``$m_{t}$'' distribution of the $Wj^{n}$ background after
hard cuts must be predicted, with small errors. We will argue later
that this is very difficult.
Consequently, we doubt that a straightforward counting experiment can yield,
on its own, a satisfactory measurement of the cross-section for the production
of single top.

\subsection{Shape Variables}

Under the assumption that a counting experiment is insufficient, we
turn to observables that (as in \cite{cdfrunone,cdfruntwo}) make use of other
aspects of the signal. In particular, we will now explore variables that take
advantage of the very special shape of single top production compared to the
background, and are less subject to, or less sensitive to, systematic errors. \ 

As noted earlier, the dominant production process is $tbq$, in which there is
a hard lepton, and also a hard
untagged jet $j$ with pseudo-rapidity\ $|\eta|\sim1-3$ and
$p_{T}$ typically larger than $25$ GeV. This strongly forward
or backward jet is a distinctive
signature which the backgrounds do not share. The $b$ jet from the $t$ decay
tends to be produced centrally (low pseudo-rapidity) with high-$p_{T}$, and is
typically the tagged jet (for which reason we only use the tagged jet in the
$t$ reconstruction.) The other $b$ jet tends to have low $p_{T}$; it is often
unobserved, and is rarely tagged \cite{ssw}.

Importantly, the $tbq$ process arises from an initial state light quark or
antiquark (typically a valence quark carrying moderate to large Bjorken-$x$)
scattering off a gluon (typically carrying lower $x$.)  This unusual
initial state has important kinematic consequences.
Because of this kinematic effect, the structure of the proton, and the
details of the electroweak theory, the $tbq$ signal has strong and
distinctive correlations and asymmetries which we can use to separate
it from the backgrounds.

First, the unusual kinematics and the structure of the proton combine
in an interesting way.  The creation of positively-charged top quarks
(and consequently positively-charged leptons in the final state) in
the $t\bar bq$ process requires either a $u$ or $\bar{d}$ in the
initial state, so that a $W^{+}$ can be emitted from the quark
line. Since a reasonably large value of Bjorken-$x$ is necessary in
order to produce the top quark, the required initial state is most
often obtained from a valence quark striking a gluon. The most likely
initial state uses a valence $u$ quark from the $p$; the next most
likely draws a valence $\bar{d}$ from the $\bar{p}$, and thereafter we
must draw on the sea quarks from either $p$ or $\bar{p}$.  Since the
quark usually has larger $x$ than the gluon that it strikes, the
$t\bar bq$ system is typically boosted in the direction of travel of
the initial quark.  Consequently, for $t$ quarks, the $t\bar bq$
system is more likely to be boosted in the proton direction than in
that of the antiproton, by a factor of roughly 2:1.  The reverse is
true for $\bar{t}$ production.  Moreover, the light-quark jet in the
final state, converted from the quark in the initial state by the
emission of the $W$, tends to travel in the proton direction when a
$t$ is produced and in the antiproton direction when a $\bar t$ is
produced.

Thus, because of
the differences between the $u$ and $d$ parton distribution functions
in the proton, and because of the quark-gluon initial state, large
asymmetries under parity P and charge conjugation C result.  These
show up strongly in the differential cross-section both for the top
(and the positively charged lepton in its decays) and for the hard
forward or backward light-quark jet.

Second, the momentum vectors of the lepton and the light-quark jet are
correlated, as a result of both kinematics and spin polarization
effects. The structure of the electroweak interactions ensures that
the spin of the top quark tends to align with the momentum direction
of the light-quark jet. Since the top decays before this spin information
is lost, it is transferred to the momentum of the final-state charged
lepton. In the $t$ rest frame the lepton momentum and jet momentum
tend to align \cite{spin}.  The boost of the top quark relative to the
lab frame, whose sign is also aligned with the momentum direction of
the jet, further tends to push both lepton and jet into the same
hemisphere.

These properties strongly distinguish the $tbq$ process from its
$t\bar{t}$ background.  At tree-level, $t\bar{t}$ is separately C and
P invariant.  Clearly this is true of the process $gg\rightarrow
t\bar{t}$, since the initial state is P-invariant on average. It is
also true of the process $q\bar{q}\rightarrow g\rightarrow t\bar{t}$,
because the intermediate gluon state is a C-eigenstate; this is the
same as in $e^{+}e^{-} \rightarrow\mu^{+}\mu^{-}$, where there is no
forward-backward asymmetry in the $\mu^{+}$ distribution. The
parity-invariance is violated at the next order, due to radiative
effects \cite{kuhnstudy}; this is a few percent effect, both small and
calculable. Moreover, there should be no strong correlation between
the momenta of the lepton and the jets in the final state.  In those
$t\bar t$ events which have both low $H_{T}$ and a reconstructable
semileptonically-decaying top quark, one or two jets from the
hadronically-decaying top tend to be lost or mismeasured. Meanwhile 
any high-$p_{T}$ untagged jets whose momenta we
might choose to compare with that of the lepton will also stem from
the hadronically-decaying top. The accidents which lead to the
selection of any given jet as part of our analysis should largely wash
out any correlation of its momentum with that of the lepton.  Indeed,
our simulations show that in $t\bar t$ the correlation between the
lepton and the highest-$p_{T}$ untagged jet (which we will
use in our analysis below) is roughly a 10\% effect.

Similar considerations apply, to a good approximation, to those QCD
events which might pass our cuts.  Parity asymmetries for these events
are small.  The charge of a fake leptons is unrelated to its momentum
direction.  Consequently, all distributions for events with fake leptons
are invariant under flipping the sign of the lepton charge;
as we will see, this imples that these events have
P-invariant distributions for the observables we will choose.
Isolated leptons from heavy flavor stem mainly from $c\bar c$ and $b
\bar b$ events; these have similar C and P properties to $t\bar t$, so
we expect small parity-asymmetries.  Meanwhile, a fake lepton and the
highest-$p_{T}$ untagged jet in the event should be essentially
uncorrelated.  However, if this jet contains heavy flavor,
then a correlation can arise when the
lepton observed in the event is from a wide-angle semileptonic
decay of a heavy-quark within the jet.  
The precise size of jet-lepton correlations from
this source is unknown to us, and is detector- and cut-dependent.  
The small overall number of QCD events
entering the single top samples at DZero \cite{d0runtwo} suggests that
QCD contributions to jet-lepton correlations are not a major issue for
the single top measurement, except possibly in the case of muons from
heavy flavor decays.  We will return to this possible exception in
section III.D.

The parity asymmetry in $Wj^{n}$ unfortunately has the same sign as
that of $tbq$, although it is less pronounced. \ The reasons for this
are easy to see. As with a $t$ quark, a $W^{+}$ is most likely to be
produced moving in the proton direction, since it is most often
produced in a $u\bar d$ event.  This leads to a well-known asymmetry
in its pseudo-rapidity. When produced in conjunction with two jets,
the $W^{+}$ still tends to be boosted in the proton direction, since
its initial state is most often $ug$ or $u\bar d$. This leads to
parity asymmetries which, though relatively small, are still quite
large in absolute size compared to the signal.\footnote{Note these
asymmetries in pseudo-rapidity, or equivalently angle, for fixed
charge, are due to the Tevatron's proton-antiproton initial state. At
the LHC, with a proton-proton initial state, the same effects will
show up as \textit{charge} asymmetries for fixed angle.} 
Unfortunately, the size of the asymmetries
and correlations in $Wj^{n}$ appears to be very sensitive to
assumptions, cuts, Monte Carlo parameters, and tagging, and will be a
source of significant systematic error. We will return to this issue
later.

\subsection{Consequences of Parity and Correlations}

In order to make the best use of these special properties of the
signals and backgrounds, it is useful to consider these issues more
formally. We next discuss the effect of C and P (non)-invariance, and
of lepton-jet pseudo-rapidity\ correlations, on two-dimensional
distributions in pseudo-rapidity.  The ``jet'' used throughout
the analysis below is always {\it the highest-$p_T$ untagged jet} in
the event.

The $p\bar{p}$ initial state of the Tevatron is a CP eigenstate, and so all
distributions of final state particles are CP-invariant (to an excellent
approximation, violated principally by the detector). This means that 
the differential cross-section $d^{2}\sigma^{+}/d\eta_{j}d\eta_{\ell}$ with
respect to the rapidities of the jet and the positively charged lepton, and
the corresponding distribution $d^{2}\sigma^{-}/d\eta_{j}d\eta_{\ell}$ for
processes with a \textit{negatively} charged lepton, \textit{must} be related
by CP:
\[
{\frac{d^{2}\sigma^{+}}{d\eta_{j}d\eta_{\ell}}}(\eta_{j},\eta_{\ell}%
)={\frac{d^{2}\sigma^{-}}{d\eta_{j}d\eta_{\ell}}}(-\eta_{j},-\eta_{\ell})\ .
\]
Consequently, we can combine data from positively and negatively charged
leptons by defining a \textit{lepton-charge-weighted pseudo-rapidity},
$\hat{\eta}_{j}=Q_{\ell}\eta_{j}$, $\hat{\eta}_{\ell}=Q_{\ell}\eta_{\ell}$,
where $Q_{\ell}$ is the lepton charge. (The variable $\hat{\eta}_{j}$ was
already introduced in \cite{cdfrunone,cdfruntwo}.) For the remainder of this
article, our entire discussion is based on the explicitly CP-invariant
differential cross section
\[
{\frac{d^{2}\sigma}{d\hat{\eta}_{j}d\hat{\eta}_{\ell}}}(\hat{\eta}_{j}%
,\hat{\eta}_{\ell})\equiv{\frac{d^{2}\sigma^{+}}{d\eta_{j}d\eta_{\ell}}}%
(\eta_{j}=\hat{\eta}_{j},\eta_{\ell}=\hat{\eta}_{\ell})+{\frac{d^{2}\sigma
^{-}}{d\eta_{j}d\eta_{\ell}}}(\eta_{j}=-\hat{\eta}_{j},\eta_{\ell}=-\hat{\eta
}_{\ell})\ .
\]

However, the $p\bar{p}$ initial state is not an eigenstate of either C or P.
Consequently, in general we expect parity-non-invariance
\[
{\frac{d^{2}\sigma^{+}}{d\eta_{j}d\eta_{\ell}}}(\eta_{j},\eta_{\ell}%
)\neq{\frac{d^{2}\sigma^{+}}{d\eta_{j}d\eta_{\ell}}}(-\eta_{j},-\eta_{\ell
})\ ,
\]
and similar non-invariance under charge conjugation
\[
{\frac{d^{2}\sigma^{+}}{d\eta_{j}d\eta_{\ell}}}(\eta_{j},\eta_{\ell}%
)\neq{\frac{d^{2}\sigma^{-}}{d\eta_{j}d\eta_{\ell}}}(\eta_{j},\eta_{\ell})\ .
\]
(Indeed these two statements are equivalent due to CP-invariance.) In terms of
the combined differential cross-section, P (and C) non-invariance implies
\[
{\frac{d^{2}\sigma}{d\hat{\eta}_{j}d\hat{\eta}_{\ell}}}(\hat{\eta}_{j}%
,\hat{\eta}_{\ell})\neq{\frac{d^{2}\sigma}{d\hat{\eta}_{j}d\hat{\eta}_{\ell}}%
}(-\hat{\eta}_{j},-\hat{\eta}_{\ell})\ .
\]

Conversely, if we were to study a \textit{parity-symmetric} process, such as
the tree-level production of $t\bar t$, it would satisfy
\[
{\frac{d^{2}\sigma}{d\hat{\eta}_{j}d\hat{\eta}_{\ell}}}(\hat{\eta}_{j}%
,\hat{\eta}_{\ell}) ={\frac{d^{2}\sigma}{d\hat{\eta}_{j}d\hat{\eta}_{\ell}}%
}(-\hat{\eta}_{j},-\hat{\eta}_{\ell})\ \ \ \ \ (\mathrm{parity-even\ process})
\ .
\]

Next, let us consider the effect of correlations. If the dynamics of a process
is such that the jet and lepton directions are uncorrelated, then the
differential cross-section factorizes into a product of two distributions, one
for the jet and one for the lepton:
\[
{\frac{d^{2}\sigma}{d\hat{\eta}_{j}d\hat{\eta}_{\ell}}}=f(\hat{\eta}%
_{j})g(\hat{\eta}_{\ell})\ \ \ \ (\mathrm{jet \ and \ lepton\ uncorrelated})
\ .
\]
Failure of this relation is proof of correlations.  These might stem
directly from correlations in the rest frame of the $tbq$ system.
However, even if the distributions in the rest frame are uncorrelated,
they will be correlated in the lab frame, once they are convolved with
a distribution of boosts of the rest frame.\footnote{For
example, suppose 
${\frac{d^{2}\sigma}{d\hat{\eta}_{j}d\hat{\eta}_{\ell}}}=
f(\hat{\eta}_{j})g(\hat{\eta}_{\ell})$ in the $tbq$
rest frame, with Gaussian lepton and jet distributions:
\[
f\left(  \widehat{\eta}_{j}\right)     
\propto
e^{-A_j\widehat{\eta}_{j}^{2}},
\quad 
g\left(  \widehat{\eta}_{\ell}\right)  \propto
e^{-A_\ell\widehat{\eta}_{\ell}^{2}}
\]
Suppose further that the rest frame is boosted by an amount
$\eta_b$ with a propability which also has a Gaussian distribution 
\[
p(\eta_b) \propto e^{-B\eta_b^2}
\]
Then the observed distribution in the lab frame is
\[
\frac{d\sigma}{d\widehat{\eta}_{j}d\widehat{\eta}_{\ell}}  
\propto%
{\displaystyle\int\limits_{-\infty}^{\infty}}
d\eta_{b}\ f\left(  \widehat{\eta}_{j}-\eta_{b}\right)  \ g\left(
\widehat{\eta}_{\ell}-\eta_{b}\right)  \ p(\eta_b)
\propto
e^{-\left[  \left(  B+A_\ell\right)A_j
\widehat{\eta}_{j}^{2}
+\left(
B+A_j
\right) A_\ell  \widehat{\eta}_{\ell}^{2}-A_jA_\ell\widehat{\eta}_{j}\widehat{\eta}_{\ell}
\right]  /\left(  A_j+A_\ell+B\right)  }.
\]
This is a correlated distribution; it cannot be written as $F(\eta_j)G(\eta_\ell)$, because
of the cross-term in the exponent.
Note the correlation vanishes in the limit of a very narrow
distribution of boosts, $B\rightarrow\infty$, even
if the boost distribution
is not centered at zero.}

Finally, if the process is both uncorrelated \textit{and} P-invariant, then
both distributions must be even functions of their particle's pseudo-rapidity.
In short
\[
{\frac{d^{2}\sigma}{d\hat{\eta}_{j}d\hat{\eta}_{\ell}}}=f(\hat{\eta}%
_{j})g(\hat{\eta}_{\ell})\ ;\ f(\hat{\eta}_{j})=f(-\hat{\eta}_{j}%
)\ ;\ g(\hat{\eta}_{\ell})=g(-\hat{\eta}_{\ell}) \ \
\ (\mathrm{uncorrelated,\ parity-even}) \ .
\]
with the four-way consequence
\begin{equation}
{\frac{d^{2}\sigma_{{}}}{d\hat{\eta}_{j}d\hat{\eta}_{\ell}}}(\hat{\eta}%
_{j},\hat{\eta}_{\ell})={\frac{d^{2}\sigma_{{}}}{d\hat{\eta}_{j}d\hat{\eta
}_{\ell}}}(-\hat{\eta}_{j},-\hat{\eta}_{\ell})={\frac{d^{2}\sigma_{{}}}%
{d\hat{\eta}_{j}d\hat{\eta}_{\ell}}}(\hat{\eta}_{j},-\hat{\eta}_{\ell}%
)={\frac{d^{2}\sigma_{{}}}{d\hat{\eta}_{j}d\hat{\eta}_{\ell}}}(-\hat{\eta}%
_{j},\hat{\eta}_{\ell}) \ \ \label{fourway}%
\end{equation}
for such processes.\footnote{Note the logic is not reversible; a
distribution satisfying the condition (\ref{fourway}) is not
necessarily uncorrelated.} The $t\bar{t}$ process does indeed satisfy
the relation (\ref{fourway}) at the 90\% level. We believe this
continues to next-to-leading order: one-loop effects cause a 5\%
parity asymmetry in $t$ and $\bar{t}$ production angles
\cite{kuhnstudy}, which, when translated into jet and lepton
pseudo-rapidities, is unlikely to violate parity by more than 10\%
(though this has not been simulated for our specific cuts.) We believe
that QCD contributions to the sample are similarly in good agreement
with (\ref{fourway}), except possibly for lepton-jet correlations in
events with heavy flavor. The $Wj^{n}$ background, with its moderate
asymmetries and correlations, accords with (\ref{fourway}) only to a
very rough approximation.  And as we have emphasized, the $tbq$ signal
strongly violates the relation (\ref{fourway}).

\subsection{Strategy and Tactics}

\begin{figure}[ptb]
\begin{center}
\includegraphics[
height=4.3016in,
width=5.5754in
]{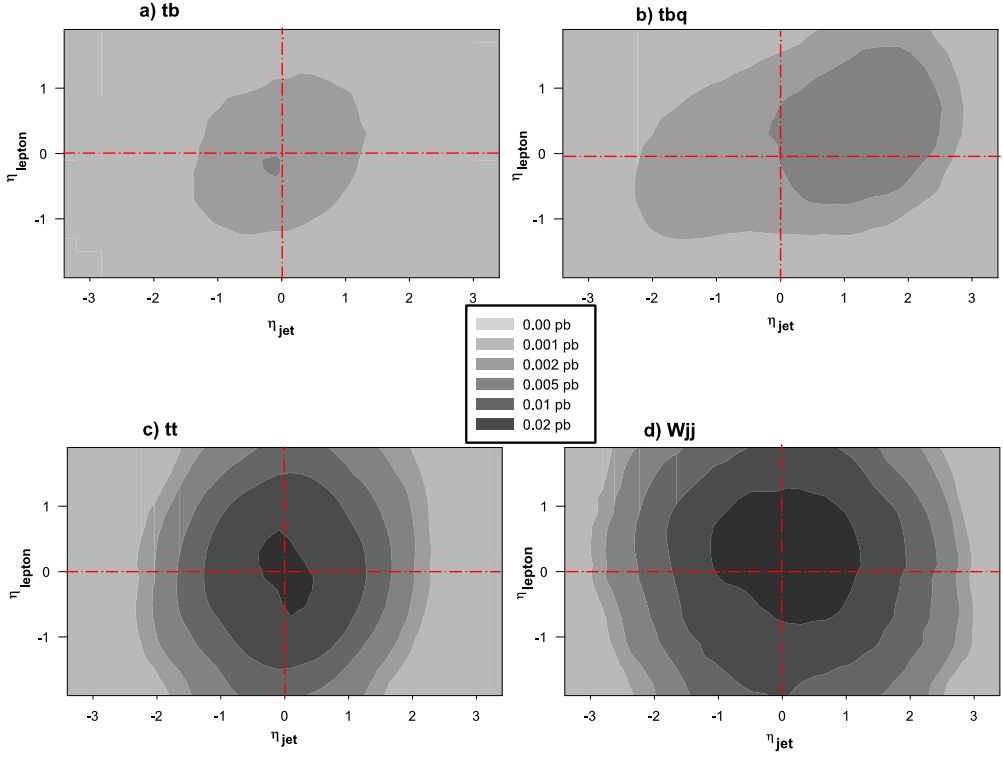}
\end{center}
\caption{Differential cross-section ($d^{2}\sigma/d\hat{\eta}_{j}d\hat{\eta
}_{\ell}$, summed over $t$ and $\overline{t}$, $e$ and $\mu$) for the a) $tb$
channel b) $tbq$ channel, c) $t\overline{t}$ channel, and d) $Wjj$ channel,
after $b$-tagging and the cuts of Table \ref{cuts3}. }%
\label{all4sig}%
\end{figure}

To illustrate the characteristic properties of the various channels, we will
study a simulated event sample defined by the cuts in Table \ref{cuts3}.
\begin{table}[ptb]
\medskip
\par
\begin{center}%
\begin{tabular}
[c]{|c|c|c|}\hline\hline
\multicolumn{1}{||c|}{Item} & \multicolumn{1}{||c|}{$p_{T}$} &
\multicolumn{1}{||c||}{$\left\vert \eta\right\vert $}\\\hline\hline
$\ell^{\pm}$ & $\geq15$ GeV & $\leq2$\\\hline
MET $\left(  \nu\right)  $ & $\geq15$ GeV & -\\\hline
jet $\left(  b\text{-tag}\right)  $ & $\geq40$ GeV & $\leq2$\\\hline
jet $\left(  \text{no }b\text{-tag}\right)  $ & $\geq30$ GeV & $\leq
3.5$\\\hline
& Min & Max\\\hline\hline
$H_{T}$ & none & $\leq300$ GeV\\\hline
\textquotedblleft$m_{t}$\textquotedblright & $\geq155$ GeV & $\leq200$
GeV\\\hline
\end{tabular}
\end{center}
\caption{\textquotedblleft Relaxed\textquotedblright\ cuts for analysis.}%
\label{cuts3}%
\end{table}These cuts are more \textquotedblleft relaxed\textquotedblright%
\ than those of Tables \ref{cuts2i} and \ref{cuts2h}, keeping a larger
fraction of both the signal and the background, and yielding a sample which we
believe is less sensitive to the systematic uncertainties stemming from the
cuts. We will argue below that shape considerations will allow us to make some
headway toward separating signal and background. The simulated differences in
shape are summarized in the contour plots of Fig.~\ref{all4sig}, which give
the distributions ($d^{2}\sigma/d\hat{\eta}_{j}d\hat{\eta}_{\ell}$) of the
various processes, plotted as functions on 
the $(\hat{\eta}_{j},\hat{\eta}_{\ell})$ plane. 
(Recall that we define the \textquotedblleft jet\textquotedblright\ of
relevance to be the highest-$p_{T}$ untagged jet.)

\begin{figure}[ptbh]
\begin{center}
\includegraphics[
height=1.9951in,
width=2.5858in
]{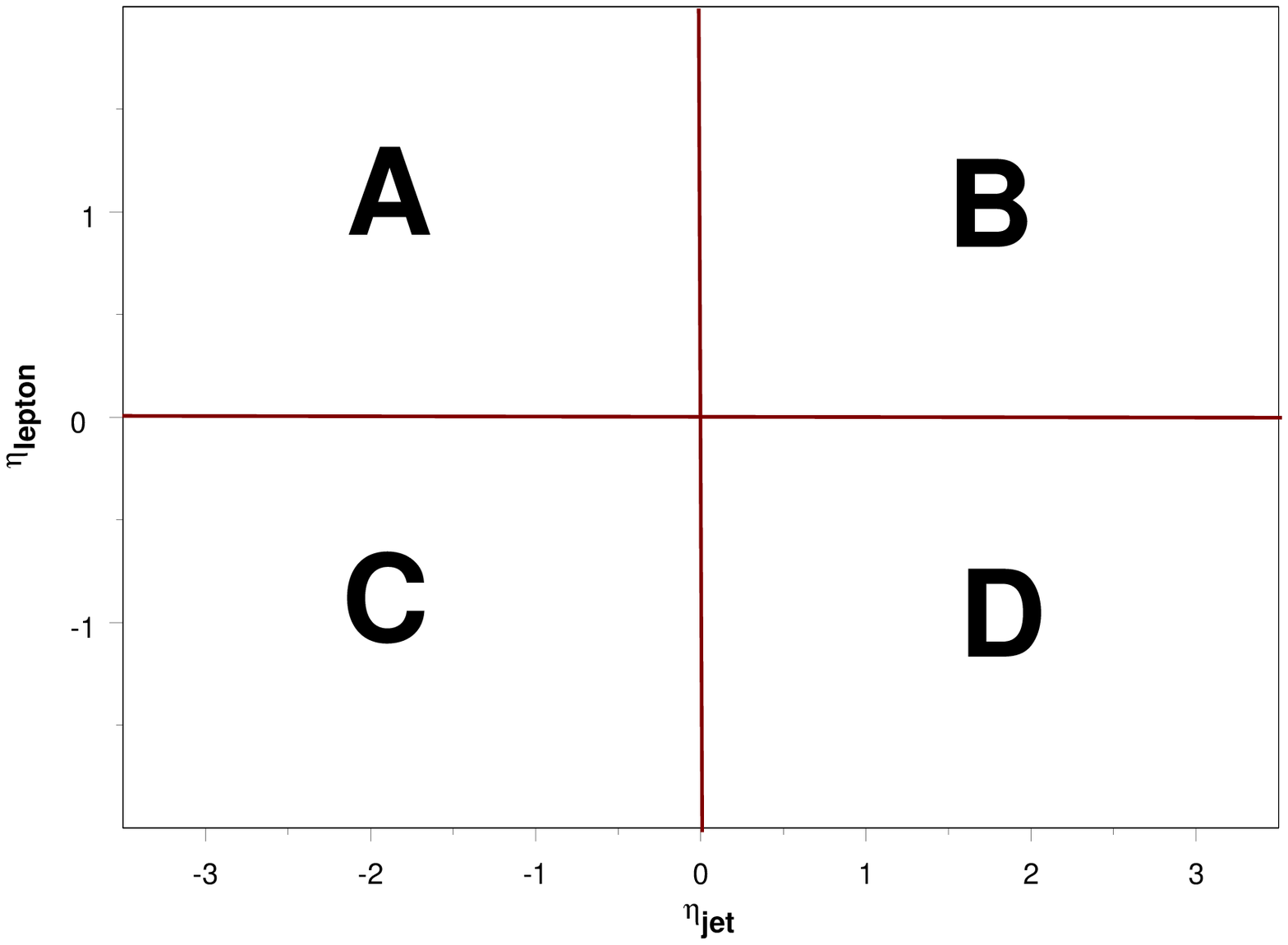}
\end{center}
\caption{The four quadrants of the $(\hat{\eta}_{j},\hat{\eta}_{\ell})$
plane.}%
\label{quads}%
\end{figure}

Figure \ref{all4sig} illustrates the degree to which the various
processes exhibit correlations and asymmetries.  We label
the four
quadrants of the $(\hat\eta_j,\hat\eta_\ell)$ plane  A, B, C,
D as shown in Fig.~\ref{quads}.  Positive jet-lepton correlations cause events
to pile up in quadrants B and C, while parity asymmetries appear in
the differences between quadrants B and C, and between quadrants A and
D.  The $tbq$ signal in Fig.~\ref{all4sig}b shows clearly shows both
effects.  The $Wj^n$ background, in Fig.~\ref{all4sig}d, also has an
asymmetric shape, though to a lesser relative degree.  The $tb$
process, Fig.~\ref{all4sig}a, shows some correlation but no asymmetry,
while Fig.~\ref{all4sig}c illustrates the uncorrelated and symmetric
nature of (tree-level) $t\bar t$.

\begin{table}[ptb]
\medskip
\par
\begin{center}%
\begin{tabular}
[c]{|c|c|c|c|c|}\hline\hline
\multicolumn{1}{||c|}{Channel} & \multicolumn{1}{||c|}{A} &
\multicolumn{1}{||c|}{B} & \multicolumn{1}{||c|}{C} &
\multicolumn{1}{||c||}{D}\\\hline\hline
$tb$ & $8$ & $11$ & $12$ & $8$\\\hline
$tbq$ & $14$ & $41$ & $18$ & $22$\\\hline
$t\bar{t}$ & $105$ & $109$ & $106$ & $105$\\\hline
$Wjj$ & $204$ & $207$ & $159$ & $180$\\\hline
\end{tabular}
\end{center}
\caption{Numbers of events for 3 fb$^{-1}$ in 
the four quadrants for various
channels (summed over $t$ and $\overline{t}$, $e$ and $\mu$). See
Fig.~{\ref{quads}} for definition and labels of the quadrants.}%
\label{quadtable}%
\end{table}

As a quantitative measure of these statements, we consider the
differential cross-sections integrated separately over the four
quadrants of the $(\hat{\eta}_{j},\hat{\eta}_{\ell})$ plane. For a
given luminosity $\mathcal{L}$, the number of $tbq$ events in the A
quadrant is
\[
\mathcal{L} \times
\int_{0}^{2}\ d\hat\eta_{\ell}\int_{-3.5}^{0}\ d\hat\eta_{j}\ {\frac
{d^{2}\sigma^{tbq}}{d\hat{\eta}_{j}d\hat{\eta}_{\ell}}}\ ,
\]
that in the B quadrant is
\[
\mathcal{L} \times
\int_{0}^{2}\ d\hat\eta_{\ell}\int_{0}^{3.5}\ d\hat\eta_{j}\ {\frac
{d^{2}\sigma^{tbq}}{d\hat{\eta}_{j}d\hat{\eta}_{\ell}}}\ ,
\]
and so forth. \ The resulting numbers of events with $\mathcal{L}$ 
= 3 fb$^{-1}$ of data
appear in Table \ref{quadtable} (summed over $t$ and $\overline{t}$, $e$ and
$\mu$). \ Statistical uncertainties in each bin are uncorrelated with other
bins. Systematic uncertainties in these numbers, which are very substantial
for $Wj^{n}$, deserve considerable discussion; since this table is merely
intended for a general illustration, we defer this discussion until Sec.~IV.

To capture quantitatively these differences in shape between signal and
background, we suggest defining three orthogonal functions in the $(\hat{\eta
}_{j},\hat{\eta}_{\ell})$ plane, based on the formal discussion of the
previous section. For any differential cross-section, whether a signal, a
background or a combination, and whether simulated or measured experimentally,
we may write it as a sum of three components
\begin{equation}
{\frac{d^{2}\sigma_{{}}}{d\hat{\eta}_{j}d\hat{\eta}_{\ell}}}(\hat{\eta}%
_{j},\hat{\eta}_{\ell})=\bar{F}(\hat{\eta}_{j},\hat{\eta}_{\ell})+F_{+}%
(\hat{\eta}_{j},\hat{\eta}_{\ell})+F_{-}(\hat{\eta}_{j},\hat{\eta}_{\ell}),
\label{sigmaFFF}%
\end{equation}
where the components are of the form%
\begin{equation}
\bar{F}(\hat{\eta}_{j},\hat{\eta}_{\ell})\equiv{\frac{1}{4}}\left[
{\frac{d^{2}\sigma_{{}}}{d\hat{\eta}_{j}d\hat{\eta}_{\ell}}}(\hat{\eta}%
_{j},\hat{\eta}_{\ell})+{\frac{d^{2}\sigma_{{}}}{d\hat{\eta}_{j}d\hat{\eta
}_{\ell}}}(-\hat{\eta}_{j},-\hat{\eta}_{\ell})+{\frac{d^{2}\sigma_{{}}}%
{d\hat{\eta}_{j}d\hat{\eta}_{\ell}}}(\hat{\eta}_{j},-\hat{\eta}_{\ell}%
)+{\frac{d^{2}\sigma_{{}}}{d\hat{\eta}_{j}d\hat{\eta}_{\ell}}}(-\hat{\eta}%
_{j},\hat{\eta}_{\ell})\right]  \label{Fbardef}%
\end{equation}%
\begin{equation}
F_{+}(\hat{\eta}_{j},\hat{\eta}_{\ell})\equiv{\frac{1}{4}}\left[  {\frac
{d^{2}\sigma_{{}}}{d\hat{\eta}_{j}d\hat{\eta}_{\ell}}}(\hat{\eta}_{j}%
,\hat{\eta}_{\ell})+{\frac{d^{2}\sigma_{{}}}{d\hat{\eta}_{j}d\hat{\eta}_{\ell
}}}(-\hat{\eta}_{j},-\hat{\eta}_{\ell})-{\frac{d^{2}\sigma_{{}}}{d\hat{\eta
}_{j}d\hat{\eta}_{\ell}}}(\hat{\eta}_{j},-\hat{\eta}_{\ell})-{\frac
{d^{2}\sigma_{{}}}{d\hat{\eta}_{j}d\hat{\eta}_{\ell}}}(-\hat{\eta}_{j}%
,\hat{\eta}_{\ell})\right]  \label{Fplusdef}%
\end{equation}%
\begin{equation}
F_{-}(\hat{\eta}_{j},\hat{\eta}_{\ell})\equiv\frac{1}{2}\left[  {\frac
{d^{2}\sigma_{{}}}{d\hat{\eta}_{j}d\hat{\eta}_{\ell}}}(\hat{\eta}_{j}%
,\hat{\eta}_{\ell})-{\frac{d^{2}\sigma_{{}}}{d\hat{\eta}_{j}d\hat{\eta}_{\ell
}}}(-\hat{\eta}_{j},-\hat{\eta}_{\ell})\right]  \label{Fminusdef}%
\end{equation}

Let us comment on some properties of these functions. First, they are explicit
functions, not abstract devices: they can be directly constructed from any
finite data set, simulated or measured. Second, they are orthogonal in the
sense that
\[
\int_{-a}^{a}d\hat{\eta}_{j}\int_{-b}^{b}d\hat{\eta_{\ell}}\ \bar{F}\ F_{\pm
}=0\ ;\int_{-a}^{a}d\hat{\eta}_{j}\int_{-b}^{b}d\hat{\eta_{\ell}}\ F_{+}%
F_{-}=0\ ;\
\]
(where $a$ and $b$ are arbitrary positive numbers); indeed this orthogonality
applies in any symmetrically-defined region of the $(\hat{\eta}_{j},\hat
{\eta_{\ell}})$ plane. Third, the functions provide important physical
information about the symmetry properties of the differential cross-section.
$\bar{F}$ and $F_{+}$ are parity-even while $F_{-}$ is parity-odd; thus
$F_{-}=0$ (within statistics) for any parity-even distribution. Meanwhile,
because of Eq.~(\ref{fourway}), $F_{+}$ will also vanish if the distribution
is parity-even \textit{and} the leptons and jets are uncorrelated. Fourth, by
construction, these functions have special symmetries under reflection in the
$(\hat\eta_{j},\hat\eta_{\ell})$ plane, as will be obvious in the figures
below. $\bar{F}$ and $F_{+}$ have four-way symmetry; in both cases, it is
sufficient to know the function in any one quadrant to know it in all four
quadrants. Meanwhile $F_{-}$ has two-way symmetry; quadrants A and D are
related, as are B and C, but quadrants A and B are independent and must be
determined separately.\footnote{In short, $F_{-}$ contains twice as much
information as $F_{+}$ and $\bar{F}$. In principle one could further separate
$F_{-}$ into two orthogonal functions, but this turns out not to be
particularly useful.}

In the problem at hand, the fact that the signal has strong asymmetries and
correlations, while the backgrounds do not, is very usefully characterized
using these functions. In particular, we expect, based on the properties we
have discussed above, that
\[
\bar{F}^{t\bar{t}}\gg |F_{\pm}^{t\bar{t}}|\ ;\ 
\bar{F}^{Wj^{n}}>|F_{\pm}^{Wj^{n}}|\ 
;\ \bar{F}^{tbq}\sim |F_{\pm}^{tbq}|\ .
\]
Our simulations further suggest that one can find cuts that are feasible at
the Tevatron such that the backgrounds are still very large but only in
$\bar{F}$, with
\[
\bar{F}^{t\bar{t}}\sim\bar{F}^{Wj^{n}}\gg\bar{F}^{tbq}\ ,
\]
while the signal has a much larger role to play in the other functions:
\[
|F_{+}^{t\bar{t}}|\sim| F_{+}^{Wj^{n}}|\sim| F_{+}^{tbq}|\ ,
\ |F_{-}^{t\bar{t}}|\ll
|F_{-}^{Wj^{n}}|\sim |F_{-}^{tbq}|\ .
\]
especially away from the center of the $(\hat{\eta}_{j},\hat{\eta}_{\ell})$
plane. (The $tb$ signal is smaller than the $tbq$ signal for all quantities,
but especially for $F_{+}$, by a factor of about 3, and for $F_{-}$, by a
factor of order 10.)

\begin{figure}[ptbh]
\begin{center}
\includegraphics[
height=4.2125in,
width=5.4613in
]{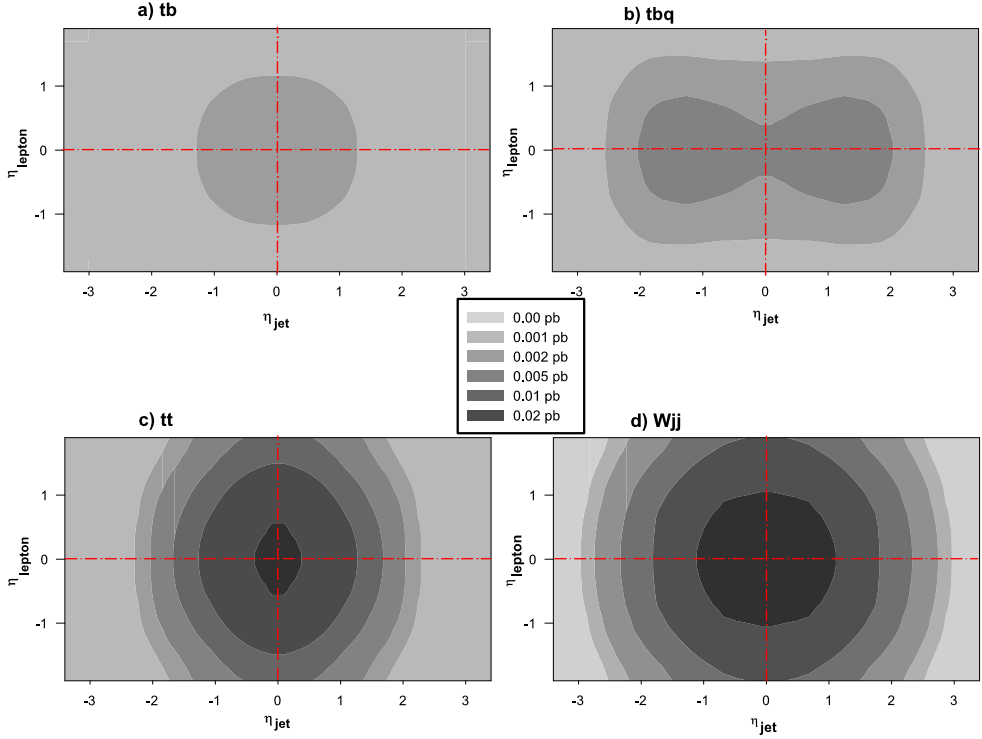}
\end{center}
\caption{Contour plots for the function $\overline{F}\left(  \hat{\eta}%
_{j},\hat{\eta}_{\ell}\right)  $ for a) $tb$, b) $tbq$, c) $t\overline{t}$,
and d) $Wjj$ channels (summed over $t$ and $\overline{t}$, $e$ and $\mu$). }%
\label{allfbar}%
\end{figure}

\begin{figure}[ptb]
\begin{center}
\includegraphics[
trim=0.000000in 0.000000in -0.106586in -0.132145in,
height=4.216in,
width=5.4198in
]{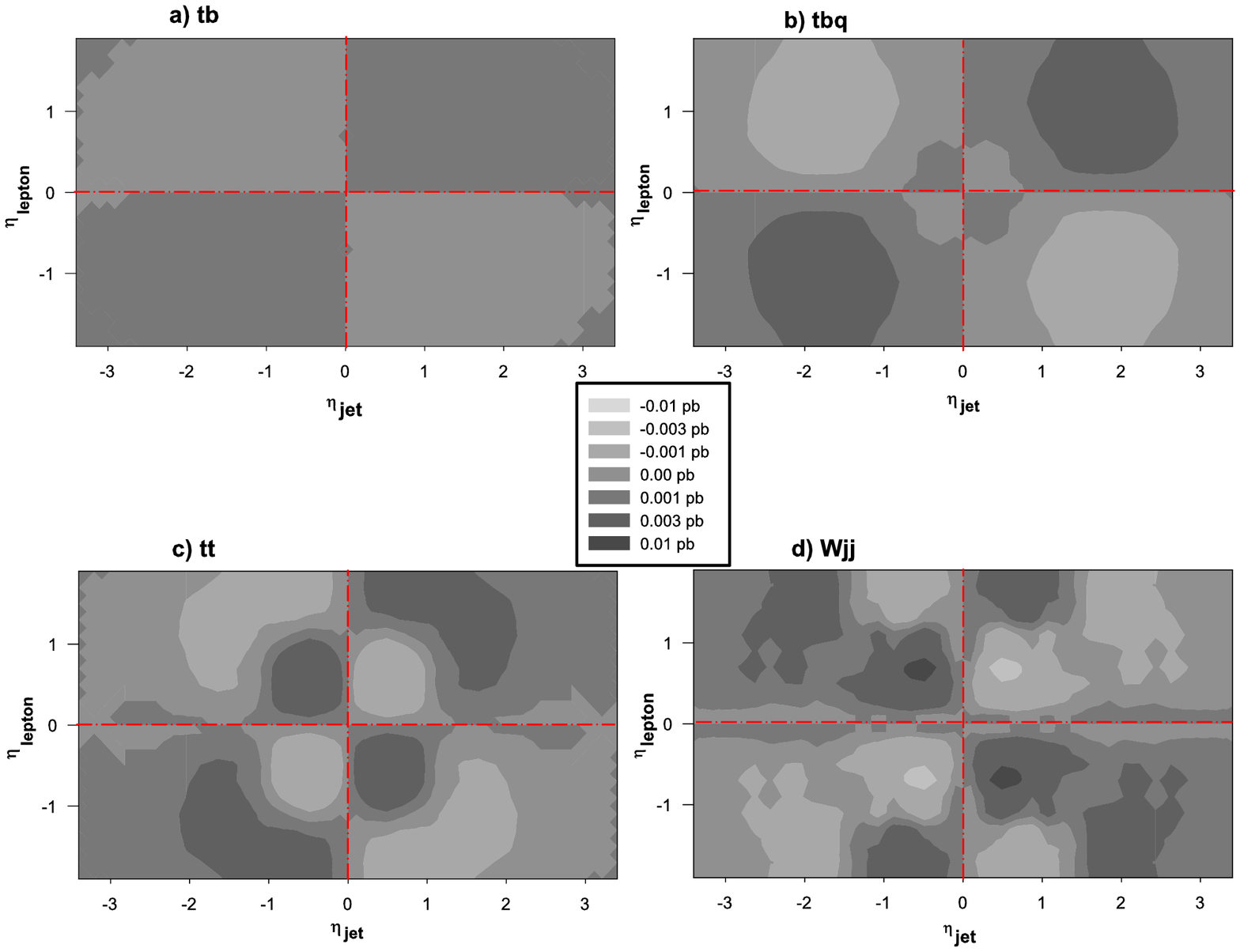}
\end{center}
\caption{Contour plots for the function $F_{+}\left(  \hat{\eta}_{j},\hat
{\eta}_{\ell}\right)  $ for a) $tb$, b) $tbq$, c) $t\overline{t}$, and d)
$Wjj$ channels (summed over $t$ and $\overline{t}$, $e$ and $\mu$). }%
\label{allfsym}%
\end{figure}

\begin{figure}[ptbh]
\begin{center}
\includegraphics[
height=4.216in,
width=5.4613in
]{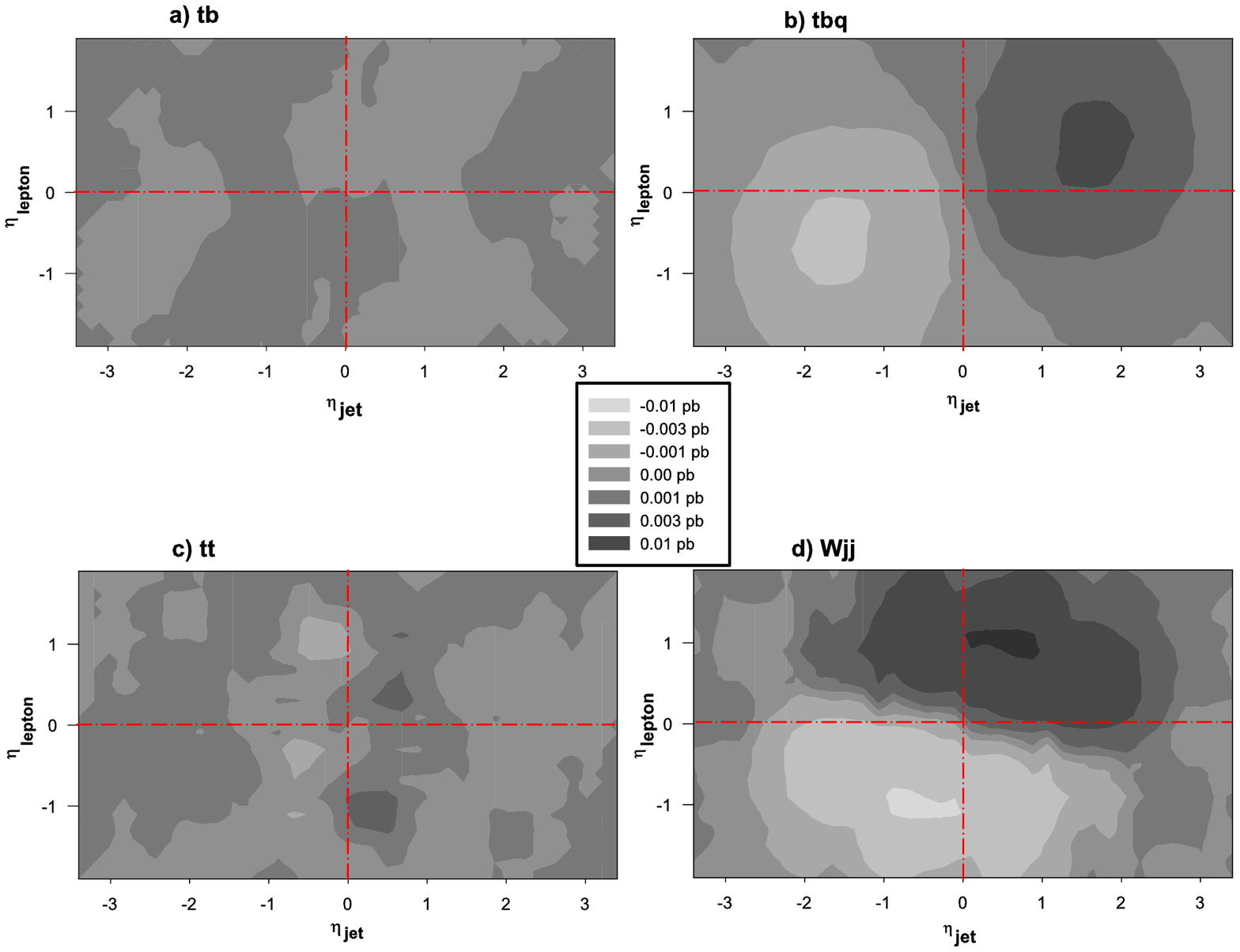}
\end{center}
\caption{Contour plots for the function $F_{-}\left(  \hat{\eta}_{j},\hat
{\eta}_{\ell}\right)  $ for a) $tb$, b) $tbq$, c) $t\overline{t}$, and d)
$Wjj$ channels (summed over $t$ and $\overline{t}$, $e$ and $\mu$). }%
\label{allfasym}%
\end{figure}

These claims are illustrated in Figs.~\ref{allfbar}, \ref{allfsym} and
\ref{allfasym}, where the functions $\bar{F},F_{+},F_{-}$ are shown, for $tb$,
$tbq$, $t\bar{t}$, and $Wj^{n}$. (The reader should note that the scale for
the contours in Figs.~\ref{allfsym} and \ref{allfasym} differs from that used
in Figs.~\ref{all4sig} and \ref{allfbar}; this is because of the smaller
dynamic range in the $F_{\pm}$ distributions.)  The symmetry
properties of the three functions discussed earlier are clearly
evident.

To make this comparison more concrete, the integrals of the 3 functions over
quadrants A and B, for an integrated luminosity $\mathcal{L}=$ 3 fb$^{-1}$, are
presented in Table \ref{Fquads}; this table can be constructed from Table
\ref{quadtable}.  The definitions of the entries in the table
are
\begin{align*}
\overline{F}_{\text{A}}  &  \equiv%
\mathcal{L}\times
{\textstyle\int\nolimits_{\text{A}}}
d\hat{\eta}_{j}\ d\hat{\eta}_{\ell}\ \overline{F}
 =\overline{F}_{\text{B}}=\overline{F}_{\text{C}}=\overline
{F}_{\text{D}},\ \\
F_{+,\text{A}}  &  \equiv%
\mathcal{L}\times
{\textstyle\int\nolimits_{\text{A}}}
d\hat{\eta}_{j}\ d\hat{\eta}_{\ell}\ F_{+}
=-F_{+,\text{B}}=-F_{+,\text{C}}=F_{+,\text{D}}\ ,\\
F_{-,\text{A}}  &  \equiv%
\mathcal{L}\times
{\textstyle\int\nolimits_{\text{A}}}
d\hat{\eta}_{j}\ d\hat{\eta}_{\ell}\  F_{-}
=-F_{-,\text{D}}\ ,\\
\quad F_{-,\text{B}}  &  \equiv%
\mathcal{L}\times
{\textstyle\int\nolimits_{\text{B}}}
d\hat{\eta}_{j}\ d\hat{\eta}_{\ell} \ F_{-}
=-F_{+,\text{C}}.
\end{align*}
The table also shows the statistical errors in these quantities. Since this
table is intended only to emphasize qualitative points, we postpone discussion
of systematic errors and next-to-leading order corrections until we outline a
more sophisticated approach with better statistical errors.

As we argued before, we are justified in disregarding QCD events.  The
number of QCD events entering the sample is small \cite{d0runtwo},
\[
\bar{F}^{t\bar{t}},\bar{F}^{Wj^{n}}\gg\bar{F}^{QCD}\sim\bar{F}^{tbq},\bar
{F}^{tb}\ .
\]
For the DZero detector, 
the number of events  in the electron channel is smaller than the
number in the muon channel.
We expect lepton-jet correlations and parity asymmetries to be
extremely small for fake leptons; for isolated leptons from heavy
flavor, parity asymmetries are very small at tree-level, and small
at higher-orders, while lepton-jet correlations
might be a bit larger.  Thus
\[
\bar{F}^{QCD} > |F_+^{QCD}| > |F_-^{QCD}| \ ,
\]
even for isolated leptons.
This leads us to expect
\[
|F_{\pm}^{tbq}|\gg |F_{\pm}^{QCD}|\ .
\]
except possibly for $F_+$ in the muon channel.  The dominant source
for $F_+$ will be from muons emitted at large angles during $b\bar b$
and $c\bar c$ events.  The kinematics of these events can be studied
using a double-tagged sample.  We believe, therefore, that even if the
contribution of QCD muon events to $F_+$ is large enough to be a
concern, its size and shape can be determined from the data.

\begin{table}[ptb]
\medskip
\par
\begin{center}%
\begin{tabular}
[c]{|c|c|c|c|c|}\hline\hline
\multicolumn{1}{||c|}{Channel} & \multicolumn{1}{||c|}{$\overline{F}%
_{\text{A}}=\overline{F}_{\text{B}}$} & \multicolumn{1}{||c|}{$F_{+,\text{B}%
}=-F_{+,\text{A}}$} & \multicolumn{1}{||c|}{$F_{-,\text{A}}$} &
\multicolumn{1}{||c||}{$F_{-,\text{B}}$}\\\hline\hline
$tb$ & $9.8\ \pm\ 1.6$ & $1.4\ \pm\ 1.6$ & $0.0\ \pm\ 2.1$ & $-0.3\ \pm
\ 2.4$\\\hline
$tbq$ & $23.8\ \pm\ 2.4$ & $5.6\ \pm\ 2.4$ & $-3.7\ \pm\ 3.0$ & $11.6\ \pm
\ 3.8$\\\hline
$t\bar{t}$ & $106.1\ \pm\ 5.2$ & $1.2\ \pm\ 5.2$ & $-0.2\ \pm\ 7.2$ &
$1.3\ \pm\ 7.3$\\\hline
$Wjj$ & $187.7\ \pm\ 6.9$ & $-4.6\ \pm\ 6.9$ & $11.8\ \pm\ 9.8$ &
$23.8\ \pm\ 9.6$\\\hline
\end{tabular}
\end{center}
\caption{Numbers of events for 3 fb$^{-1}$ in $\overline{F}$, $F_{+}$ and
$F_{-}$, integrated over quadrants A and B, for the various channels (summed
over $t$ and $\overline{t}$, $e$ and $\mu$). Errors shown are statistical
only. See text for further interpretation.}%
\label{Fquads}%
\end{table}

\subsection{Lessons and Caveats}

We can now extract some important lessons from Table \ref{Fquads}. Before
doing so, we should comment on its limitations.

First, our study is done entirely using tree-level short-distance matrix
elements; only the normalizations are at next-to-leading order. We must
therefore emphasize that \textit{the figures and tables in this paper are
meant for illustration only.} Next-to-leading-order corrections
to the matrix elements will change the shapes of the background, in ways which
contribute nontrivially to Table \ref{Fquads}. For example, we expect the
above-mentioned next-to-leading-order effects on $t \bar t$ to affect our
estimates of $F_{\pm}^{t\bar{t}}$, by something of order $10\%$ of $\bar
{F}^{t\bar{t}}$, which is not negligible. Such changes are large enough that
they must be calculated and/or measured, but
are small enough that they do not invalidate the \textit{methodology} we are
outlining here. 

Second, it is clear from the table that statistical errors from the background
are large in $F_{+}$ and $F_{-}$, comparable to the signal. While this
looks discouraging, it applies for the distributions integrated over the
entire A or B quadrant. A quick examination of Fig.~\ref{all4sig}
shows that the situation is not quite as bad as it appears,
if one considers the region away from the center of the $(\hat\eta_{j}%
,\hat\eta_{\ell})$ plane, where the backgrounds are much smaller and the
signals are still quite large. We will discuss this in much more detail in the
next section, where we will do a more sophisticated analysis, but for the
moment we simply note that the size of the statistical errors is misleadingly
large in the above table. Still, we will see that the situation with
statistics remains unsatisfactory.

With these caveats in mind, we return to Table \ref{Fquads}, on the
basis of which we suggest the following general approach. One should
first construct, for both the data and the Monte Carlo simulation
output, the $\bar{F}$, $F_{+}$ and $F_{-}$ functions. Using these
functions, as well as information obtained from other measurements,
one can systematically test one's understanding of each process. \ The
$\bar F$ function allows a measurement of the sum of the backgrounds
without much contamination from signal. We assume that the separation
of $Wj^{n}$ from $t\bar t$ can be obtained using the fact that the
$t\bar t$ process can be measured and predicted with reasonable
accuracy, using other data samples and Monte Carlo simulation. One can
then cross-check one's understanding of the shape of the $Wj^{n}$
background using the part of the $F_{-}$ distribution (located roughly
in quadrant A) where the signal is negligible.  Finally, one can
attempt to measure the signal from $F_{+}$ and from a different part
(located largely in quadrant B) of the $F_{-}$
distribution.\footnote{Note that the study in \cite{cdfruntwo}
measures half this information; it is able to measure $\bar F$ and the
difference of quadrants $A$ and $B$ in $F_{-}$.}  Effects on $F_+$
from QCD events with isolated leptons 
can be cross-checked by study of double-tagged events,
and by comparing $F_+$ for electrons against $F_+$ for muons.

We now proceed to refine this approach, and to estimate the
associated uncertainties.

\section{Uncertainties and Optimization}

Our goal in this section is to show that the measurement of the signal
using $F_{+}$ and $F_{-}$ is potentially feasible, though difficult.
We begin this section with an overview that, using
Figs.~\ref{sigbkallbw} and \ref{sigrootall} as guides, lays out our
main points.  We then turn to more detailed consideration of
statistical and systematic uncertainties, especially those associated
with $W$-plus-jets.

\subsection{Overview}

\begin{figure}[ptbh]
\begin{center}
\includegraphics[
height=4.03688in,
width=6.29263in
]{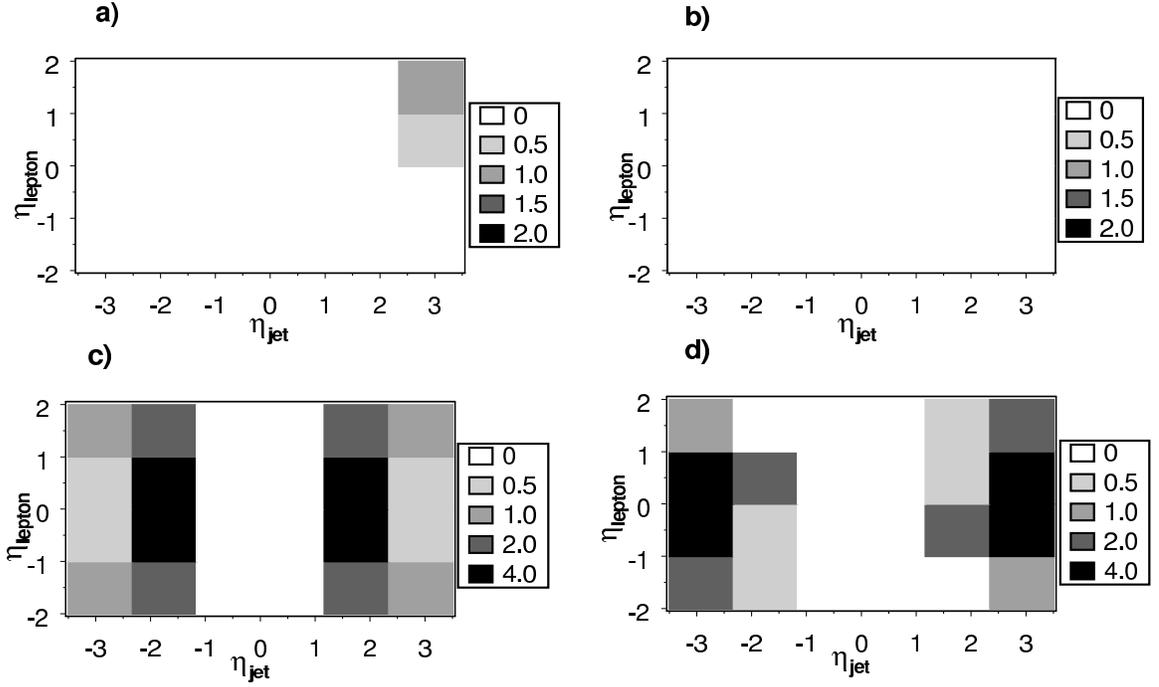}
\end{center}
\caption{Signal-to-background ratio $S/B$ across the $(\hat\eta_{j},\hat
\eta_{\ell})$ plane, showing, within each $1.16\times1.0$ bin in
$\widehat{\eta}_{j}\times\widehat{\eta}_{\ell}$, the ratio of signal to
background for a) ${\displaystyle\int}d^{2}\sigma/d\widehat{\eta}_{j}%
d\widehat{\eta}_{\ell}$, b) $\overline{F}$, c) $|F_{+}|$ and d) $|F_{-}|$.}%
\label{sigbkallbw}%
\end{figure}

Figure \ref{sigbkallbw} shows the signal-to-background ratio for the
differential cross-sections and for the three orthogonal functions that we
have defined. Figure \ref{sigbkallbw}b shows the ratio
\[
{\frac{\bar{F}^{tbq}+\bar{F}^{tb}}{\bar{F}^{t\bar{t}}+\bar{F}^{Wjj}}}%
\]
according to our simulations. Nowhere in the plane is the signal-to-background
ratio of order unity, implying that sensitivity to even small systematic
errors is severe. Instead, $\bar{F}$ should be viewed as insensitive to the
signal, and therefore mainly useful in helping determine of the size of the backgrounds.

Figures \ref{sigbkallbw}c and \ref{sigbkallbw}d show the analogous ratios
\[
\Bigg|{\frac{F_{+}^{tbq}+F_{+}^{tb}}{F_{+}^{t\bar{t}}+F_{+}^{Wjj}}}\Bigg|
\]
and
\[
\Bigg|{\frac{F_{-}^{tbq}+F_{-}^{tb}}{F_{-}^{t\bar{t}}+F_{-}^{Wjj}}}\Bigg|
\]
(We use absolute values for these two functions, as both numerator and
denominator can be negative.) For both $F_{+}$ and $F_{-}$, the
signal-to-background ratio reaches unity in some parts of the plane, implying
that, for sufficiently high statistics, \textit{the signal can be measured in
these two observables even if the background has relatively large systematic
errors.} In both cases, the dark regions are the best ones for the
measurement; the other regions should be cut away.

\begin{figure}[ptbh]
\begin{center}
\includegraphics[
height=4.0284in,
width=6.1756in
]{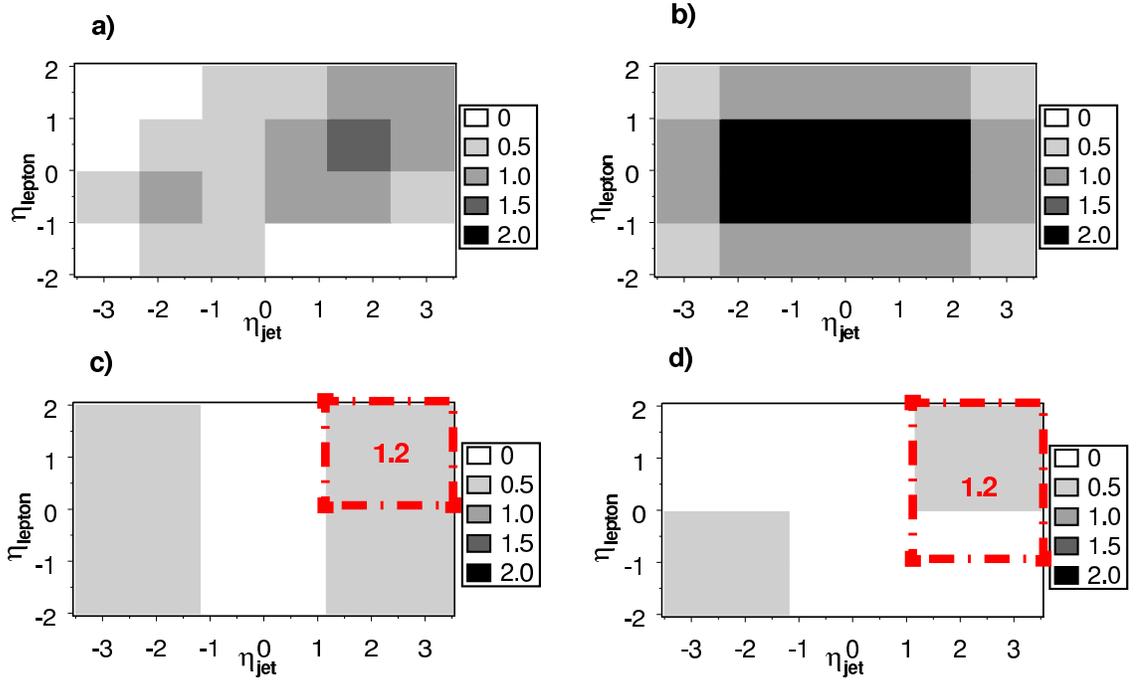}
\end{center}
\caption{Statistical significance $S/\sqrt{B+S}$, for an integrated luminosity
of 3 fb$^{-1}$, across the $(\hat\eta_{j},\hat\eta_{\ell})$ plane. Within each
$1.16\times1.0$ bin in $\widehat{\eta}_{j}\times\widehat{\eta}_{\ell}$, the
number of events in a) ${\displaystyle\int}d^{2}\sigma/d\widehat{\eta}%
_{j}d\widehat{\eta}_{\ell}$, b) $\overline{F}$, c) $|F_{+}|$ and d) $|F_{-}|$,
divided by the appropriate square-root of the number of events in background
plus signal (see text for formulas.) For each box outlined in dot-dashed
lines, the accompanying number indicates the improved value of $S/\sqrt{B+S}$
that results when the bins inside the box are combined. }%
\label{sigrootall}%
\end{figure}

In Fig.~\ref{sigrootall} the ratio of signal to the square-root of
background-plus-signal is plotted, for a integrated luminosity of 3
$\mathrm{fb}^{-1}$. Figure \ref{sigrootall}b shows
\[
{\frac{\bar{F}^{tbq}+\bar{F}^{tb}}{\frac{1}{2}{\left(  \bar{F}^{t\bar{t}}%
+\bar{F}^{Wjj}+\bar{F}^{tbq}+\bar{F}^{tb}\right)  ^{1/2}}}}%
\]
while Figs.~\ref{sigrootall}c and \ref{sigrootall}d show
\[
{\frac{|F_{+}^{tbq}+F_{+}^{tb}|}{\frac{1}{2}{\left(  \bar{F}^{t\bar{t}}%
+\bar{F}^{Wjj}+\bar{F}^{tbq}+\bar{F}^{tb}\right)  ^{1/2}}}}%
\]
and
\[
{\frac{|F_{-}^{tbq}+F_{-}^{tb}|}{\frac{1}{\sqrt{2}} {\left(  \bar{F}^{t\bar
{t}}+\bar{F}^{Wjj} +\bar{F}^{tbq}+\bar{F}^{tb} + F_{+}^{t\bar{t}}+F_{+}^{Wjj}
+F_{+}^{tbq}+F_{+}^{tb}\right)  ^{1/2}}}}%
\]
(Note that, by construction, $\bar F > F_{+}$ for any distribution, so the
expression under the square-root is always positive.) The form of the
denominators (and the factors of $1/2$ and $1/\sqrt{2}$) follow from ordinary
Gaussian statistics for each bin, using the definitions of $\bar{F}$ and
$F_{\pm}$, Eqs. (\ref{Fbardef})--(\ref{Fplusdef}), as linear combinations of
statistically independent bins.\footnote{For instance, $F_{-}$, defined in
Eq.~(\ref{Fminusdef}), satisfies
\[
\Delta F_{-} = \frac12 \left[  \left\{ \Delta\left( \frac{d^{2}\sigma}%
{d\hat{\eta}_{j}d\hat{\eta}_{\ell}}(\hat\eta_{j}, \hat\eta_{\ell})\right)
\right\} ^{2} + \left\{ \Delta\left( \frac{d^{2}\sigma}{d\hat{\eta}_{j}%
d\hat{\eta}_{\ell}}(-\hat{\eta}_{j},-\hat{\eta}_{\ell})\right) \right\} ^{2}
\right] ^{1/2} \ .
\]
Since the two terms in the right-hand side are uncorrelated,
\[
\Delta F_{-} = \frac12 \left[  \left( \frac{d^{2}\sigma}{d\hat{\eta}_{j}%
d\hat{\eta}_{\ell}}(\hat{\eta}_{j},\hat{\eta}_{\ell})\right)  +  \left(
\frac{d^{2}\sigma}{d\hat{\eta}_{j}d\hat{\eta}_{\ell}}(-\hat{\eta}_{j}%
,-\hat{\eta}_{\ell})\right) \right] ^{1/2} = {\frac{1}{\sqrt{2}}} \left( \bar F +
F_{+}\right) ^{1/2} \ ,
\]
where we used Eqs.~(\ref{Fbardef}) and (\ref{Fplusdef}). This explains the
formula used in Fig.~\ref{sigrootall}d.} The possibility of aggregating bins
into larger regions, in which these ratios are of order 1.2, is illustrated
using the dot-dash outlined bins in the case of $F_{+}$ and $F_{-}$.

Figures~\ref{sigbkallbw} and \ref{sigrootall} show that \textit{there is
nonzero overlap, for both $F_{+}$ and $F_{-}$, between the region where the
statistics is best and the region where sensitivity to systematic errors is
small. } Moreover, we learn that \textit{the two measurements should use data
predominantly from certain ``windows'' in the $(\hat\eta_{j},\hat\eta_{\ell})$
plane.} The region where the jet has small $|\hat\eta_{j}|$ has severe
problems with statistical background and sensitivity to systematics. The
$F_{+}$ measurement should be made, roughly speaking, by using all data for
$\hat\eta_{j}>1$ and any $\hat\eta_{\ell}$, (recall that the regions with either
or both $\hat\eta$ negative are redundant in the case of $F_{+}$.) Meanwhile
the $F_{-}$ measurement of the signal should be made, roughly, by cutting away
all but the region $\hat\eta_{j}>1$ and $\hat\eta_{\ell}> -1$ (the
mirror-symmetric region $\hat\eta_{j}<-1$ and $\hat\eta_{\ell}< 1$ being
redundant in this case.)\footnote{We do not mean to imply that we are
specifying the precise form of these windows. Their shapes must be optimized
for each particular analysis, based on the associated backgrounds,
acceptances, detector issues, and integrated luminosity, as well as improved
simulations. Also, no window cut need actually be used; a fit or neural
network will naturally weight the data inside the windows heavily, while
down-weighting the regions outside the windows.}

Let us also call attention to the region $\hat\eta_{j}<1$ and $\hat\eta_{\ell
}>1$ for $F_{-}$ (ignoring the redundant mirror-symmetric region), away from
the center of the plane but also outside the above-mentioned $F_{-}$ window.
Here, the signal is very small. This region is useful for a check of the
modeling of the $Wj^{n}$ background. In the standard model, the contribution
of $F_{-}$ should be consistent with the $Wj^{n}$ background, with little
$t\bar t$ and QCD background, and no measurable signal. Verifying that this is
the case is both a cross-check on the analysis and a test of the standard
model, so the measurement of $F_{-}$ in this region is also very important.

\subsection{Statistical Errors}

To explore the size of statistical errors, we have carried out two
unsophisticated likelihood analyses, one optimistic, one overly pessimistic.
In both cases, we assume 3 fb$^{-1}$ of integrated luminosity. We assume that
the shapes of the backgrounds and signals are known, and we fit for their
normalizations. This is not quite appropriate, since some information about
the normalization of the backgrounds will be known from other sources, and
since there are significant uncertainties in the shape functions, especially
for $Wj^{n}$, but we believe the measure we obtain is not far from the truth,
and that the correct lesson can be extracted.

In the first likelihood analysis, somewhat optimistic, it is assumed that the
shape of the distribution, in the $(\hat\eta_{j},\hat\eta_{\ell})$ plane, of
the sum of all backgrounds, $t\bar t$ plus $Wj^{n}$ (plus QCD) is known.
Similarly, it is assumed that the shape of the signal, $tbq$ plus $tb$, is
known. Maximizing the likelihood as a function of the relative normalization
of the signal over background, we find that the measurement of signal over
background can be made to a precision of about 40 percent.

For the second likelihood analysis, it is assumed that the shape of $t\bar t$
(plus QCD) is known, the shape of $Wj^{n}$ is separately known, and, as
before, the shape of the signal $tbq$ plus $tb$ is known. We then fit for the
two relative normalizations, and find that the measurement of the signal over
the background can be made to a precision of about 50 percent. Meanwhile the
measurement of the relative normalizations of the two backgrounds can be
performed with an uncertainty of order 15 percent.

Both of these analyses show that statistical uncertainties are large
even for $\mathcal{L}$ = 3 fb$^{-1}$, probably comparable to or larger
than the theoretical uncertainties in the shapes of the distributions
for the signals and backgrounds. To the extent that normalizations of
the backgrounds can be pinned down using other information, the
situation may be slightly better than this estimate suggests.

Intuitively, from the figures, the region in the center of the $(\hat\eta
_{j},\hat\eta_{\ell})$ plane plays a large role in fixing the backgrounds,
while the large-$\hat\eta_{j}$ region of quadrant B plays a large role in
fixing the signal. Indeed, as noted earlier, the statistical significance of
the measurement in the signal comes dominantly from the regions outlined in
dot-dashed lines in Fig.~\ref{sigrootall}. Both measurements of $F_{+}$ and
$F_{-}$ in these ``windows'' have $S/\sqrt{B+S}$ of order 1.2, reasonably
consistent with the above likelihood analyses, which were applied to the full
distributions over the entire plane.

\subsection{Systematic Errors: General Comments}

We now turn our focus to systematic uncertainties, which mainly stem from an
inability to predict and simulate the backgrounds. Our use of the $F_{-}$ and
$F_{+}$ functions, we will argue, helps us reduce the sensitivity of the
measurement to systematic uncertainties. However, the systematic errors on the
backgrounds are very large at present, and must still be reduced.

Earlier in this section, using Fig. \ref{sigbkallbw}, we discussed the
signal-to-background ratios for the various component functions. We noted that
Fig.~\ref{sigbkallbw}b indicates that $\bar F$ has a poor signal-to-background
ratio. Although it has been argued that systematic uncertainties in predicting
$\bar F$, using a combination of Monte Carlo simulations and data, will not be
large, even a 15 percent systematic uncertainty is already enough to make
$\bar F$ problematic for measuring the signal, no matter how good the
statistics. Instead, it is better to treat $\bar F$ as a measurement which,
along with other inputs, is used to help fix the normalizations of the
backgrounds.\footnote{In the counting experiment discussed in Sec.~III.A, one
essentially uses $\bar F$ to make the measurement, though with much tighter
cuts than the ``relaxed cuts'' employed here.}

Fig.~\ref{sigbkallbw}c and \ref{sigbkallbw}d show that the situation
with $F_{\pm}$ is much better, outside of the central region of the
plane where the background peaks. $F_{+}$ may receive some additional
contributions which we have neglected. There is a QCD contribution,
which we have argued is probably small, especially in the electron
channel. Detector effects and correlations from cuts can also
contribute to $F_{+}$. This means the signal-to-background ratio in
the figure may be an overestimate.\footnote{Moreover, it appears in
the figure to be somewhat better than we actually expect it to be,
perhaps by as much as a factor of two in any given bin. In our
simulations there is some accidental cancellation between the $Wj^{n}$
and $t\bar t$ contributions to $F_{+}$; see
Fig.~\ref{allfsym}. Unfortunately our knowledge of these two
backgrounds is too poor to be certain that this cancellation is
robust, and moreover statistical fluctuations may also ruin the
cancellation.}

Meanwhile, $F_{-}$, the P-odd C-odd observable, has the feature that many
potential sources of systematic uncertainty largely cancel. We expect QCD
backgrounds, correlations from cuts, and effects of detector cracks or damage
to be small in this variable, since they are largely P-even and/or C-even.
Many uncertainties in simulating $t\bar t$ and even, to a degree, $Wj^{n}$
will have substantial cancellations. This makes this variable especially compelling.

Eventually, the contribution of $t\bar t$ to $F_{\pm}$ should not have
large systematic uncertainties.  It is important to keep in mind that
there are non-negligible effects, including a parity asymmetry
\cite{kuhnstudy}, that appear at one loop.  Next-to-leading-order
calculations of the distributions in the
$(\hat\eta_{j},\hat\eta_{\ell})$ plane are needed.  These can be
computed with small errors and should allow a precise background
subtraction of $t\bar t$.  At present, however, their absence causes a
substantial uncertainty in this subtraction.

For both $F_{+}$ and $F_{-}$, the most serious systematic problems stem from
the uncertainties in the shape of $Wj^{n}$. 
For this reason, we will now present an extended discussion of
this background. 

\subsection{Systematic Errors: Predicting $W$-plus-jets}

One might hope that systematic errors stemming from the $Wj^{n}$
background could be greatly reduced, for the variables $F_\pm$, using
a sideband subtraction in the variable ``$m_{t}$''.  This approach,
analogous to the method suggested in \cite{ssw} for use in a counting
experiment, would allow a background subtraction without much
theoretical input.  However, any such attempt will run
into the same issues that cause this method to fail for a counting
experiment.  In Fig.~\ref{fig:mtdistrib}, we showed that the
application of aggressive cuts to bring $Wj^{n}$ under statistical
control simultaneously makes a sideband analysis problematic by
deforming the shape of the $Wj^{n}$ background, leaving it
non-monotonic across the region needed for the sideband analysis.  To
do a subtraction therefore requires that the shape of the background,
after cuts, to be accurately \textit{predicted}, using a combination
of theory, Monte Carlo and data. Unfortunately, prediction of any
aspects of $Wj^{n}$, especially with one or more $b$-tagged jets, is
very difficult indeed.  As we will now discuss in detail, we believe
that Monte Carlo results for the $Wj^{n}$ sample with $b$-tags cannot
currently be trusted at the level that is likely to be needed.

The shape of the $Wj^{n}$ background is plagued with the usual concerns about
the inability of Pythia or Herwig \cite{HERWIG}
to reliably generate the correct pattern of
additional radiated jets; for recent discussion, see \cite{sullivan}. But to
this and other typical problems, which are known to be issues in many
processes, we must add some others which are specific to the sample with one
(and only one) $b$-tagged jet.  (Note the event samples
used in our single-top analysis above include events with one {\it or
more} tagged jets; however, the problems detailed in this
section are somewhat less severe
for samples with more than one tag.)

\begin{table}[ptb]
\medskip
\par
\begin{center}%
\begin{tabular}
[c]{|c|c|c|c|}\hline\hline
\multicolumn{1}{||c|}{$Wjj$ Channel} & \multicolumn{1}{||c|}{$\sigma$(before
tags) [fb]} & \multicolumn{1}{||c|}{$\sigma$(after tags) [fb]} &
\multicolumn{1}{||c||}{Fraction tagged}\\\hline\hline
$Wqq$ & $16,470$ & $192$ & $1\%$\\\hline
$Wqg$ & $32,000$ & $732$ & $2\%$\\\hline
$Wgg$ & $14,760$ & $484$ & $3\%$\\\hline
$Wcq$ & $3200$ & $318$ & $10\%$\\\hline
$Wcg$ & $2240$ & $238$ & $11\%$\\\hline
$Wc\overline{c}$ & $600$ & $104$ & $17\%$\\\hline
$Wb\overline{b}$ & $496$ & $224$ & $45\%$\\\hline\hline
Total & $69766$ & $2291$ & $3\%$\\\hline
\end{tabular}
\end{center}
\caption{The cross sections for $Wj^{n}$ events with at least two jets, before
and after tagging of one of the jets.
%
Each row refers to the
combination of processes with the tree-level final state shown in the
left-most column.  The
$Wcq$ and $Wcg$ entries include both $c$ and $\overline{c}$ quarks;
the symbol $q$ stands for $u,d,s,\bar u,\bar d,\bar s$. 
After showering, hadronization and jet identification,
and application of the cuts in Table \ref{cuts1}, the resulting cross section
is indicated in the next column. The cross section corresponding to a single
tagged jet is shown in the third column, while the last column shows the ratio
of the previous two. Simulation uncertainties in these numbers are of order
$\pm$1--3\%, except for the $Wqg$ and $Wqq$ channels, with uncertainties of
order $\pm$4\% and $\pm$8\%, respectively. 
The (much larger) systematic
uncertainties are discussed in the text.}%
\label{tagfrac}%
\end{table}

\begin{table}[ptb]
\medskip
\par
\begin{center}%
\begin{tabular}
[c]{|c|c|c|c|c|}\hline\hline
\multicolumn{1}{||c|}{$Wjj$ Channel} & \multicolumn{1}{||c|}{$b$-jet} &
\multicolumn{1}{||c|}{$c$-jet} & \multicolumn{1}{||c|}{non-$b/c$-jet} &
\multicolumn{1}{||c||}{Total}\\\hline\hline
$Wqq$ & $2\%$ & $1\%$ & $6\%$ & $9\%$\\\hline
$Wqg$ & $11\%$ & $8\%$ & $14\%$ & $33\%$\\\hline
$Wgg$ & $7\%$ & $5\%$ & $5\%$ & $17\%$\\\hline
$Wcq$ & $0\%$ & $14\%$ & $1\%$ & $15\%$\\\hline
$Wcg$ & $1\%$ & $10\%$ & $0\%$ & $11\%$\\\hline
$Wc\overline{c}$ & $0\%$ & $5\%$ & $0\%$ & $5\%$\\\hline
$Wb\overline{b}$ & $10\%$ & $0\%$ & $0\%$ & $10\%$\\\hline\hline
Total & $31\%$ & $43\%$ & $26\%$ & $100\%$\\\hline
\end{tabular}
\end{center}
\caption{A budget of the sample of $Wj^{n}$ events with a single
tagged jet (and containing at least one untagged jet), constructed
as in the previous table.
Each entry
shows the fraction of the sample containing a single-tagged
jet that was generated from the
tree-level process labelling the row (as in Table~\ref{tagfrac}) and
in which the tagged jet was of the class labelling the column. The
$Wcq$ and $Wcg$ entries include both $c$ and $\overline{c}$ quarks;
the symbol $q$ stands for $u,d,s,\bar u,\bar d,\bar s$. The entries in
this table are subject to an additive uncertainty of $\pm1-2\%$. \ The
(much larger) systematic uncertainties are discussed in the text.}%
\label{budget}%
\end{table}

A striking feature of this sample is that only a moderate fraction of
$Wj^{n}$ events with a single tagged jet (and passing our cuts)
actually have a short-distance $b$-quark parton present in the
hard-scattering process; those that do are mainly $Wb\bar{b}$. \ 
Instead, the single
$b$-tag sample is composed of many different contributions. \ This
situation is made explicit in Table \ref{tagfrac}, which shows the
cross-section for various $Wjj$ channels, before and after the
requirement of a single tag, and Table \ref{budget}, which lists the
relative contributions from the various $Wjj$ channels to the single
tagged sample. \ As with all our simulations, these contributions were
calculated using Madgraph \cite{madevent} to evaluate the parton cross
sections, Pythia \cite{pythia} to provide showering and hadronization,
and PGS \cite{pgs} as the detector simulation of jet identification
and tagging.  The total $Wj^n$ cross-section (before tagging) 
was normalized to the one-loop result \cite{MCFM}.
The basic cuts of Table \ref{cuts1} were used; note that
an untagged jet was also required in the event. 
The labels $Wxy$ in
the first column indicate the perturbative final state at tree-level
(as evaluated in Madgraph); here $q$ represents a light quark or
antiquark ($u,d,s, \bar u,\bar d,\bar s$). 
In Table \ref{tagfrac}, the
cross-section for each channel, before and after the requirement of a
single tagged jet, is shown; also shown is the  ratio of after-tagging to
before-tagging ({\it i.e.}, the fraction of each
channel containing a
single tagged jet.) In Table \ref{budget}, the three central columns
divide the tagged events by whether the jet that was tagged in the event
contained a bottom hadron, a charm hadron, or no heavy flavor. For
example, the entry in the $b$-jet column of the $Wqg$ row indicates
that 11 percent of the entire $Wj^{n}$ single-tag sample arises from
$Wqg$ events ($q=u,d,s,\bar{u},\bar{d},\bar{s})$ in which the tagged
jet contains bottom hadrons, mainly due to the splitting $g\to
b\bar{b}$ in the parton showering.  \ In determining the entries in
Table \ref{budget} from those in Table \ref{tagfrac}, we have
attempted to correct for any double counting. \ For example, we have
subtracted the contribution to the $b$-jet $Wgg$ entry (from $g\to
b\bar{b}$ in the parton shower) arising from events with kinematic
configurations already present in the $b$-jet $Wb\overline{b}$
entry. \ This leads to small corrections, of order the stated $\pm2$\%
uncertainty, in several of the entries.

The central feature of these tables is the multitude of contributions
to the tagged sample, all of similar magnitude. \ As clearly visible
in Table \ref{tagfrac}, the large parton-level cross-sections for the
light quark and gluon processes are reduced by a low tagging fraction,
while the much smaller tree-level heavy-quark cross-sections are
subject to much larger tagging rates. \ Consequently, within the
single-tag sample, all such partonic processes end up contributing at
roughly at the same level, as is clear in the ``after-tagging'' column
of Table \ref{tagfrac}, and in the ``total'' column of Table
\ref{budget}. The breakdown of the resulting single-tag sample by the
fraction of the sample for which the tagged jet is a bottom jet, a
charm jet, or a jet without heavy flavor is visible in Table
\ref{budget}; as the numbers in the bottom row indicate, all
contributions are again of the same order.

The simulations were performed with statistics sufficient to ensure such that
each entry in Table \ref{budget} is subject to an \textit{additive}
statistical uncertainty of order 2--3\%. \ However, the systematic
uncertainties in the simulations are much larger, due to a host of important
physical and technical issues. \ Since these systematic uncertainties are the
most important obstacle to an accurate background estimate, we now discuss
them in detail.

Consider first the uncertainties in the basic event simulation.\ A precision
simulation of the $Wjj$ sample cannot, at present, be carried out. The
parton-level theoretical computation of the differential cross-section for $W$
plus two or more high $p_{T}$ jets has been advanced in recent years: $Wjj$
has been calculated to next-to-leading order, and $Wjjj$ is known at
leading-order. While the program \textquotedblleft MCFM\textquotedblright%
\ \cite{MCFM,campbell} can provide accurate
next-to-leading-order parton-level cross-sections, no event generator valid at
next-to-leading order currently exists. Consequently, there is at present no
possibility of simulating this background without significant theoretical
uncertainties. \ 

Moreover, even when a next-to-leading-order event generator becomes
available, there are serious questions concerning showering algorithms
that must be addressed. The main problem is associated with the
splitting of gluons to heavy quarks (or more precisely, with tuning
Pythia or Herwig to simulate correctly the process in which a partonic
gluon generates one or more jets containing charm or bottom mesons.) \
As illustrated in Table \ref{budget}, a substantial fraction of the
single-tagged sample, of order 33\% in our simulations, originates
from this process. \ It is not known with confidence how often gluons
at short distance lead to jets with charm or bottom hadrons, or how
often this process leads to two jets rather than one. Studies on this
issue that compare data from LEP \cite{gtobb} and Tevtron \cite{CDFb}
with QCD expectations \cite{QCDgtobb} and with Pythia and Herwig
suggest that there is no serious disagreement.  However, this
conclusion rests on the substantial uncertainties (of order 30\%) in
both the experimental and the theoretical results.  Generally the
perturbative predictions (including resummed logarithms) and the Monte
Carlo results (with default parameters) tend to underestimate the
observed rates of heavy flavor production in parton showers.  Also,
the rough agreement speaks mainly to overall rates of heavy-flavor
production, not to charm-to-bottom ratios or kinematic distributions.
Given the large sensitivity of the single-tag sample to these effects,
it appears that a sizeable uncertainty in both the normalization and
shape of the $Wj^{n}$ background arises from this source.

A related issue is the role of uncertainties in parton distribution functions.
The single tag sample receives significant contributions from events with
charm in the final state that arise from initial states containing non-valence
partons. Examples include processes such as $u\bar{s}\rightarrow W^{+}u\bar
{c}$, $g\bar{d}\rightarrow W^{+}g\bar{c}$, and $\bar{u}\bar{d}\rightarrow
W^{+}\bar{u}\bar{c}$. These processes have varying shapes and rates, and
depend on the poorly-determined parton distribution functions.

Even if one could precisely simulate these events, there is still the
issue of determining the efficiencies for tagging of
jets with bottom or charm quarks, and for mistagging of jets with neither. Thus
must be done as a function of $p_{T}$ and pseudo-rapidity. As
indicated in Table \ref{budget}, each of these tagging processes plays
a comparable role in determining the $Wjj$ sample.  \ Uncertainties in
tagging functions thus lead to uncertainties in the shape of $Wj^{n}$
which cannot be ignored. An especially serious issue is that the ratio
of $c$ to $b$ tagging rates is currently extracted not from data but
from Monte Carlo programs, which, among other problems, are dependent
upon the correct modelling of gluons splitting to heavy flavor.

 Clearly, it would be
beneficial to decrease the mistagging and charm-tagging efficiencies
\textit{relative} to the efficiency for tagging of bottom jets. This would
improve the ratio of the signal to the $Wj^{n}$ background, reducing
sensitivity to systematic errors, and also would directly suppress some of the
main sources of uncertainty in the prediction of $Wj^{n}$. However, tuning the
$b$-tagging algorithm to improve purity of the sample generally comes at the
cost of a small reduction in the $b$-tagging efficiency, which in turn
decreases the signal and increases statistical errors. The right balance
between these competing issues is detector- and luminosity-dependent and must
be left to the experimental collaborations.

Another related concern is the subtle linkage between $b$-tagging, jet
definitions, and gluon splitting. When a gluon splits to two heavy quarks, the
probabilities to obtain one tagged jet, two tagged jets, or one tagged and one
untagged jet depend upon all three issues. This may mean
that the separation of the samples with one tag versus two tags is unstable
and not well predicted by theory. It is for this reason that we have chosen to
consider samples with one-or-more tagged jets. Alternatively, one might
require strong angular separation between tagged jets in order to retain predictivity.

A further issue involves the potentially large sensitivity of the
shape and normalization of the multi-channel $Wj^{n} $ background to
the cuts used to bring backgrounds under control. \ In particular,
both the $H_{T}$ and $m_{t}$ cuts used in the current analysis and in
recent experimental papers \cite{cdfruntwo} reshape the $Wj^{n}$
background. While it has been shown \cite{ssw} that jet vetoes are
effective at substantially reducing the size of $t\bar{t}$
backgrounds, we would argue against the use of this approach. The
systematic errors which jet vetoes introduce into the prediction of
the $Wj^{n}$ background has not been quantified, but we expect it is
prohibitively large. We are especially concerned about the requirement
of two-and-only-two jets employed in \cite{cdfruntwo}. We believe this
will make the prediction of $Wj^{n}$ unreliable, both because of
problems with QCD corrections, and because of failures to correctly
model the rate at which gluons in parton-level processes lead to zero,
one or two jets containing heavy flavor.  There is at present no
consensus as to the safest method for reducing $t\bar{t}$, or, for
that matter, $Wj^{n}$. We would like to argue strongly that this is a
very important problem, which our methods simultaneously mitigate and
highlight. On the one hand, neither the $F_{+}$ nor the $F_{-}$
distribution is strongly sensitive to the overall size of $t \bar{t}$.
Consequently, the use of a strong jet veto or harsh $H_{T}$ cut is
unnecessary, and indeed unjustified to the extent systematic errors in
$Wj^{n}$ become larger as a result. Our approach would instead prefer
a method which cuts $t\bar{t}$ less severely, and in a safer fashion,
such that theoretical errors in predicting $Wj^{n}$ remain small. The
$H_{T}$ cut that we use is, we believe, safer, being a cut on a more
inclusive variable; whereas the splitting of one jet into two, due to
a fluctuation or a changed jet algorithm, affects a jet veto in a
dangerous way, this is not so for an $H_{T}$ cut. But this safety is
only relative; there are problems with jets moving above or below the
minimum $p_{T}$ required for a jet to be included in the variable
$H_{T}$ that we defined. 
In any case, our method requires less focus on
reducing the size of $t\bar t$ and more focus on keeping $Wj^{n}$ as
small and as predictable as possible.

With all these problems in view, any method used to extract the $Wj^{n}$
background will have to pass many cross-checks. One important consistency
check can be carried out by calibrating Monte Carlo simulations using both
$Wj^{n}$ and the process $Z$-plus-jets (\textquotedblleft$Zj^{n}%
$\textquotedblright), where the $Z$ decays leptonically. The processes
$Wj^{n}$ and $Zj^{n}$ do not have the same shapes, rates, and heavy flavor
content, so one cannot directly take ratios of distributions or even of
overall cross-sections. However, the overall kinematics of $Zj^{n}$ is similar
to $Wj^{n}$, and is similarly sensitive to all of the above-mentioned issues.
For these reasons, we expect that matching a Monte Carlo to the rates and
shapes of the $Zj^{n}$ and $Wj^{n}$ distributions from data, with zero, one or
two tagged jets, will significantly reduce systematic uncertainties. Such
matching will require adjusting parameters which affect the splitting of
gluons to heavy flavor, and adjusting tagging functions.\footnote{We believe
there should be enough $Z$-plus-jets data, with 3 fb$^{-1}$ of integrated
luminosity, for this analysis to be carried out.} It would also be very
helpful if direct measurements of the bottom-content, charm-content, and
non-$b$/non-$c$ content of the various single- and double-tagged $Wj^{n}$ and
$Zj^{n}$ samples could be carried out, even with low precision and confidence.
This would allow direct tests of numerous Monte Carlo predictions that at
present have very large uncertainties. 
Studies of the $Zb$
to $Zj$ ratio have been carried out \cite{D0ZbZj}, but the challenging study
of $Zc$ separately should also be considered as statistics improve.

In summary, determining the single-tagged $Wj^{n}$ cross-section will require a
carefully crafted combination of theory, data, and theory-optimized Monte
Carlo.

\section{Summary}

A counting experiment for discovery of electroweak single-top production
appears very challenging. In an effort to improve the situation, we have
explored the possibility of using the distinctive shape of this process to
separate it from background. We use as observables the pseudo-rapidity
$\eta_{j}$ of the leading-$p_{T}$ untagged jet and the pseudo-rapidity
$\eta_{\ell}$ of the charged lepton, weighted by the charge of the lepton
$Q_{\ell}$. (One of these variables was already used in 
\cite{cdfrunone,cdfruntwo}.)
Considering the distributions of signal and background in the $(\hat\eta
_{j},\hat\eta_{\ell})$ plane (where $\hat\eta_{j}=Q_{\ell}\eta_{j}$ and
$\hat\eta_{\ell}=Q_{\ell}\eta_{\ell}$), we have noted that the distributions
for $t\bar t$ and for QCD are largely symmetric, while that of the signal is
not; the $Wj^{n}$ ($W$-plus-jets) background is intermediate between them.
Constructing functions $\bar F, F_{\pm}$, defined in Eqs.~(\ref{Fbardef}%
)-(\ref{Fminusdef}), which have various symmetry properties, we have shown
that the statistical and systematic errors in the functions $F_{\pm}$, which
are orthogonal to the function $\bar F$ that would be used in a counting
experiment, can be brought close to reasonable size without using extreme cuts.

Here is a summary of key ingredients that went into this analysis, as well as
a list of elements which we did not account for, and a few of our crucial assumptions:

\begin{itemize}
\item All cross-sections ($tb, tbq, t\bar t, Wbb, Wcc, Wcq, Wcg, Wqq, Wqg,
Wgg$) were calculated at tree level.

\item These were then normalized to theoretical calculations at
next-to-leading order. For $Wj^{n}$, only the sum of all channels was
normalized in this way.

\item All processes were run through Pythia, to simulate showering, and
through the detector simulation PGS.

\item We imposed hard $p_T$ cuts on the leading tagged and untagged
jets, and required that a top quark be reconstructable from the
lepton, tagged jet, and missing energy.  We applied an $H_{T}$ cut
aimed at reducing $t\bar t$; we did not use a jet veto.

\item $p_{T}$-dependence of tagging fractions was accounted for, with the
maximal tagging rates at high $p_{T}$ for $b, c, q/g$ being taken as $50\%,
15\%, 1\%$. The details of the analysis are  sensitive to these
numbers, as well as to the rate for hard gluons at leading order to evolve
into jets containing heavy flavor.

\item We did not attempt to simulate QCD events. Instead, we argued
QCD effects are (in theory) sufficiently symmetric in shape and (from
DZero data) sufficiently small in rate that their contributions to all
useful observables can be neglected (with the possible exception,
for DZero, of $F_+$ in the muon channel.)

\item We discussed important shape effects on $t\bar t$ at next-to-leading
order \cite{kuhnstudy}, which we expect to be about ten percent or less, but
did not simulate them.
\end{itemize}

It should be noted that we have not optimized our cuts to improve efficiencies
and reduce systematics in a rigorous way. Indeed, it would not be too useful
to do so, since the optimization is a moving target, depending on integrated
luminosity, on tagging rates and other detector details, on
next-to-leading-order shapes, and on Monte Carlo assumptions. We believe,
therefore, that our results could be improved upon through such an
optimization, though this is only worth doing in a concrete analysis. We also
have not explored whether other methods of reconstructing $m_{t}$ might be
more effective, or whether variables other than $H_{T}$ might be better as far
as both statistics and systematics. This is certainly something that should be
done as the integrated luminosity increases.

Moreover, there is a natural extension of our method which we did not
consider, but which should be explored if the integrated luminosity
becomes sufficiently high.  Our observable $F_+$ focussed on
lepton-jet psuedo-rapidity correlations, but as we pointed out, these
correlations can have two sources: {\it inherent} correlations in the
rest frame of the $tbq$ system (or of the top quark itself) and
correlations which are {\it induced} by the boosting of these frames
into the lab frame.  Both of these effects are present in the signal,
and they add coherently to give a large
contribution to $F_+$.  One could imagine measuring the two effects
separately.  This could potentially allow for even greater separation
of signal from background.

We conclude with a summary of and comments upon what we see as the main
lessons of our analysis.

\ 

\noindent1) Our method largely removes $t\bar t$ and QCD events from
the observables $F_\pm$, making extreme cuts on $t \bar t$
unnecessary, and focussing attention on $Wj^{n}$ as the main
background.  While the statistical fluctuations from $t\bar t$ are
still important, they are of less concern than systematic errors on
$Wj^{n}$, since the former scale as the square-root of the $t\bar t$
rate, while the latter scale linearly with the $Wj^{n}$
rate. Moreover, the $t\bar{t}$ background is much more safely
calculated and simulated, and will be, in the end, easier to
remove. In our approach to single top, \textit{one should not cut hard
on $t\bar t$ if doing so causes the systematic uncertainties in
$Wj^{n}$ to increase substantially}.  In particular, this argues
against the use of a severe jet veto, in which events with more than two
observed jets are discarded.

\ 

\noindent2) The method that we have introduced requires that the
shapes of $t\bar t$ and $Wj^{n}$ be properly modeled, but it also
provides for cross-checks. The function $\bar F$ measures the total
background, and, assuming $t\bar t$ can be determined using other
samples, this allows a measurement of the total $Wj^n$ background.
Meanwhile, there is a region in the $(\hat\eta_{j},\hat\eta_{\ell})$
plane where the $F_{-}$ function gets small contributions from both
signal and from $t\bar t$. This region allows a check of whether the
shape of $Wj^{n}$ has been correctly understood, as well as being
interpretable as a worthwhile test of the standard model itself.

\ 

\noindent3) The thorniest problem obstructing the measurement of single top is
understanding the shape of the $Wj^{n}$ background.
This is a large and irreducible background which must be subtracted from the
signal, even in the context of the methods we proposed here. This subtraction
could be done directly, using our cuts, but this requires some prediction of
the shape in the $(\hat\eta_{j},\hat\eta_{\ell})$ plane. Alternatively, a
sideband analysis around the ``$m_{t}$'' window cut could be applied to
$F_{\pm}$, in appropriate pseudo-rapidity windows, but this too requires
prediction of the effect of cuts on the  distribution of $Wj^{n}$ in ``$m_{t}%
$'' and pseudo-rapidity.

While theory, Monte Carlo and data all can, and must, assist with
these subtractions, many different types of uncertainties plague the
sample with a single $b$-tag (and therefore the sample with
one-or-more $b$-tags), making it unclear how to bring all the
available resources together. We believe that a dedicated study,
examining the rates, shapes, and flavor content (especially of bottom
versus charm) of both $Wj^{n}$ and $Zj^{n}$, with zero, one and two
tagged jets, will be necessary. This will require a blend of multiple
measurements, theoretically precise predictions, and careful tuning
and cross-checking of Monte Carlo simulations. Since this issue
affects many other measurements, including the Higgs search and
numerous beyond-the-standard-model searches, we view this as a very
high priority.

\ 

\noindent4) There are very few paths toward reducing the $Wj^{n}$ background
relative to the signal. One clear need is to decrease the mistagging rate and
charm-tagging efficiency while maintaining or increasing the $b$-tagging
efficiency; this would both improve signal-to-background and reduce some of
the uncertainties that make it difficult to model the background. Another
important step would be taken if the resolution in reconstructing the top
quark mass from the $b$, lepton and missing energy could be improved. This
would allow a narrowing of the $m_{t}$ window cut, which would both reduce the
size of the $Wj^{n}$ background and reduce experimental and theoretical
uncertainties in any sideband analysis. Beyond this, one would need to
consider more radical ideas, such as finding methods which could, on average,
differentiate bottom jets from charm jets, bottom jets from antibottom jets,
and/or jets formed by short-distance heavy quarks from jets formed by gluons
that have split into roughly-collinear heavy quark pairs.

\ 

In conclusion, our method for extracting single-top confirms that one can use
the distinctive shape of the signal to reduce backgrounds more effectively
than in a pure counting experiment. However, we also find that backgrounds are
much worse than was once thought. Improvements in (mis)tagging rates and in
the understanding thereof, careful modeling of $W$-plus-jets cross-checked
against both theory and data, and more theoretically trustworthy techniques
for cutting away backgrounds will all be necessary for a robust measurement of
single top production. The single-tag $W$-plus-jets background, in particular,
represents a challenge that the whole community must meet head-on.

\begin{acknowledgements}
We thank our colleagues G. Watts, A. Garcia-Bellido, T. Gadfort,
A. Haas, H.  Lubatti, and T. Burnett for many useful conversations and
direct assistance.  We also thank K. Ellis, T. Junk, S. Mrenna, T. Stelzer, 
Z. Sullivan and E. Thomson for extended conversations and for their thoughtful
comments and criticisms.  This work was supported by U.S. Department
of Energy grants DE-FG02-96ER40956 and DOE-FG02-95ER40893, and by an
award from the Alfred P. Sloan Foundation.
\end{acknowledgements}

\end{document}